%% file: __sample-sigconf.tex
\documentclass[sigconf]{acmart}
\usepackage{url}            
\usepackage{booktabs}       
\usepackage{amsfonts}       
\usepackage{nicefrac}       
\usepackage{microtype}      
\usepackage{bm}
\usepackage{graphicx}
\usepackage{subfigure}

\usepackage{amssymb}
\usepackage{amsmath}
\usepackage{amsthm}
\usepackage[ruled,vlined]{algorithm2e}
\usepackage{array}
\usepackage{tabularx}
\usepackage{multirow}
\usepackage{verbatim}

\usepackage{color}
\newcommand{\partitle}[1]{\vspace{2mm}\noindent \textbf{#1.}}
\usepackage{booktabs} 
\usepackage[Symbol]{upgreek}
\usepackage{stfloats}
\usepackage{diagbox}
\usepackage{natbib}
\usepackage{dsfont}

\setcopyright{rightsretained}

\begin{document}
\copyrightyear{2020}
\acmYear{2020}
\setcopyright{acmcopyright}\acmConference[KDD '20]{Proceedings of the 26th ACM SIGKDD Conference on Knowledge Discovery and Data Mining}{August 23--27, 2020}{Virtual Event, CA, USA}
\acmBooktitle{Proceedings of the 26th ACM SIGKDD Conference on Knowledge Discovery and Data Mining (KDD '20), August 23--27, 2020, Virtual Event, CA, USA}
\acmPrice{15.00}
\acmDOI{10.1145/3394486.3403264}
\acmISBN{978-1-4503-7998-4/20/08}
\fancyhead{}

\title{Semi-supervised Collaborative Filtering by Text-enhanced Domain Adaptation}

\author{Wenhui Yu}
\thanks{School of Software, Tsinghua National Laboratory for Information Science and Technology}
\affiliation{%
  \institution{Tsinghua University}
  \city{Beijing}
  \state{China}
}
\email{yuwh16@mails.tsinghua.edu.cn}

\author{Xiao Lin}
\affiliation{%
  \institution{Alibaba Group}
  \city{Beijing}
  \state{China}
}
\email{hc.lx@alibaba-inc.com}

\author{Junfeng Ge}
\affiliation{%
	\institution{Alibaba Group}
	\city{Beijing}
	\state{China}
}
\email{beili.gjf@alibaba-inc.com}

\author{Wenwu Ou}
\affiliation{%
  \institution{Alibaba Group}
  \city{Beijing}
  \state{China}
}
\email{santong.oww@taobao.com}

\author{Zheng Qin}
\authornote{The corresponding author.}
\affiliation{%
  \institution{Tsinghua University}
  \city{Beijing}
  \state{China}
}
\email{qingzh@mail.tsinghua.edu.cn}


\begin{abstract}
	Data sparsity is an inherent challenge in the recommender systems, where most of the data is collected from the implicit feedbacks of users. This causes two difficulties in designing effective algorithms: first, the majority of users only have a few interactions with the system and there is no enough data for learning; second, there are no negative samples in the implicit feedbacks and it is a common practice to perform negative sampling to generate negative samples. However, this leads to a consequence that many potential positive samples are mislabeled as negative ones and data sparsity would exacerbate the mislabeling problem. 
	
	To solve these difficulties, we regard the problem of recommendation on sparse implicit feedbacks as a semi-supervised learning task, and explore domain adaption to solve it. We transfer the knowledge learned from dense data to sparse data and we focus on the most challenging case — there is no user or item overlap.
	
	In this extreme case, aligning embeddings of two datasets directly is rather sub-optimal since the two latent spaces encode very different information. As such, we adopt domain-invariant textual features as the anchor points to align the latent spaces. To align the embeddings, we extract the textual features for each user and item and feed them into a domain classifier with the embeddings of users and items. The embeddings are trained to puzzle the classifier and textual features are fixed as anchor points. By domain adaptation, the distribution pattern in the source domain is transferred to the target domain. As the target part can be supervised by domain adaptation, we abandon negative sampling in target dataset to avoid label noise.
	
	We adopt three pairs of real-world datasets to validate the effectiveness
	of our transfer strategy. Results show that our models outperform existing models significantly.

\end{abstract}

%
%



\maketitle

\input{__samplebody-conf}

\bibliographystyle{ACM-Reference-Format}
\input{__sample-sigconf1.bbl}


\end{document}

%% file: __samplebody-conf.tex




\section{Introduction}
\label{sec:introduction}
Recommender systems gain extensive research attention due to the general use on various online platforms like E-commerce and video websites. In real-world applications, implicit feedback data (one-class data like clicks and purchases) is widely used since it is easy to collect and generally applicable. Recommendation in real-world applications usually suffers from a serious sparsity issue, which leads to two difficulties: (1) There are not enough interactions to provide information for model learning. Moreover, the data is highly unbalanced: the majority of users and items only have a few interactions with the system, which makes the recommendation tasks more difficult. (2) Since we only observe a small part of positive samples in implicit feedbacks, existing negative sampling strategies treat unobserved samples as negative ones \cite{BPR,WBPR,GBPR,SPLR,NBPO}. However, in this way, many potential positive samples are mislabeled as negative ones and the model is seriously misled by the label noise, especially on sparse data.  

To tackle the aforementioned challenges, we adopt transfer learning to enrich the information on the sparse dataset and we focus on cross-domain recommendation without user and item overlap. To be specific, we train the source and target models on the dense and sparse data respectively and explore domain adaptation \cite{domain_adaption1,domain_adaption2} to align the embeddings (i.e., latent factors), which is the key component to encode user preferences in Collaborative Filtering (CF) models. We transfer the knowledge (such as distribution patterns) learned from the dense data to the sparse data to learn better embeddings on it. Considering the target model is supervised by both the positive samples and the domain adaptation mechanism, we abandon negative sampling on the sparse data to avoid the label noise issue. As we can see, learning the target model is a semi-supervised learning task.

To illustrate our strategy, we first give a brief introduction of domain adaptation. \citet{domain_adaption2} proposed Domain Adversarial Neural Network (DANN) for unsupervised image classification tasks. In DANN, a feature extractor is trained to extract visual features and a domain classifier is trained to classify which domain the current features come from. By adversarially training the feature extractor to puzzle the domain classifier, visual features of the two domains are aligned and the knowledge learned from the source domain is transferred to the target domain. More details of domain adaptation can be found in Subsection \ref{subsubsec:domain_adaptation}.

\vspace{-2mm}
\begin{figure}[ht]
	\centering
	\subfigure[Visual features aligning in the visual space]{
		\includegraphics[scale = 0.27]{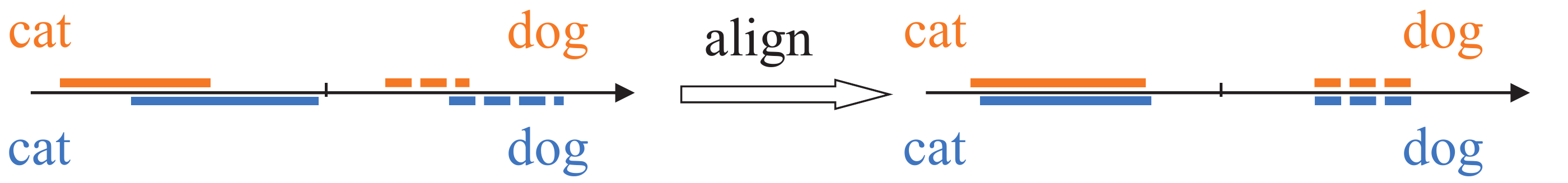}
		\label{subfig:visual_align}
	}
	\subfigure[Embedding aligning in the latent space]{
		\includegraphics[scale = 0.27]{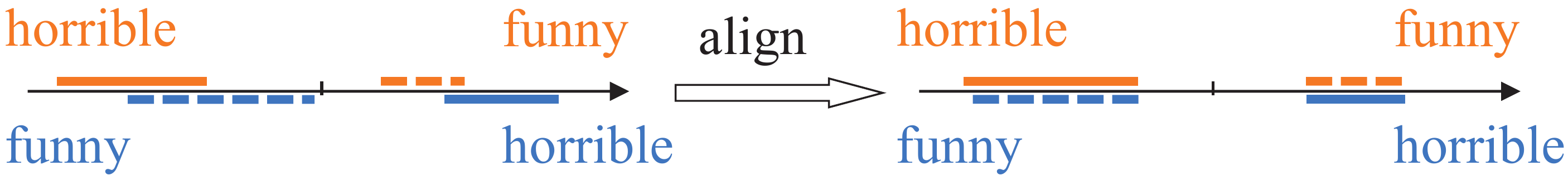}
		\label{subfig:embedding_align}
	}
	\subfigure[Embedding aligning with textual features in the latent-textual space]{
		\includegraphics[scale = 0.27]{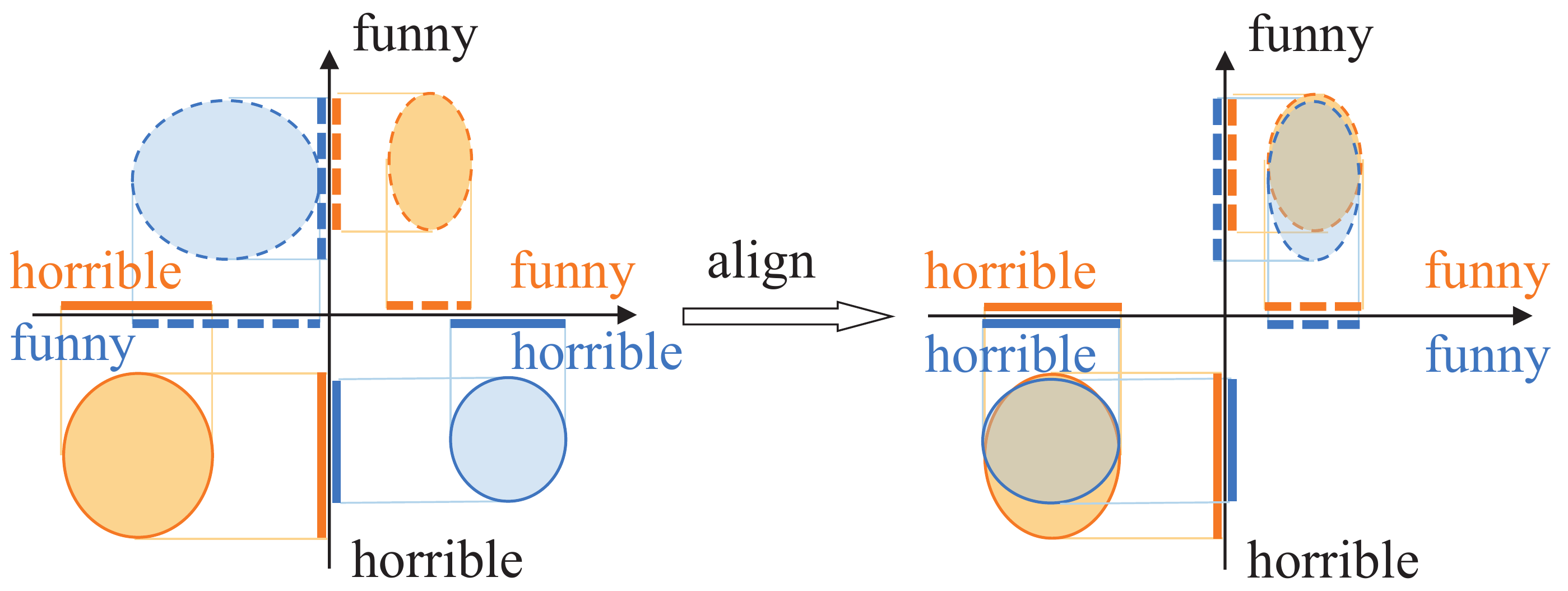}
		\label{subfig:embedding_text_align}
	}
	\caption{
		Examples of domain adaptation in computer vision tasks and recommendation tasks. We use different colors to distinguish two domains and use solid and dotted lines to distinguish different categories. For concise illustration, the visual, latent, and textual spaces are all represented as 1 dimensional spaces.}
	\label{fig:embedding_align}
\end{figure}

In this paper, we aim to align embeddings by domain adaptation in recommendation tasks, however face a critical issue. We illustrate both vanilla DANN and our strategy in Figure \ref{fig:embedding_align}. DANN is proposed for image classification tasks, which aligns high-level image representations in the visual space. As the two domains share the same feature extractor, images from the two domains are mapped into the same space, thus images with similar semantics are distributed in the similar position of the space. By domain adaptation, clusters with similar semantics are aligned together, and distribution patterns are transferred to refine the representations on the target domain. Taking the visual space represented in Figure \ref{subfig:visual_align} as an example, the negative and positive semi-axes encodes cats and dogs respectively, thus cats from different domain are both mapped to negative semi-axis and aligned together by domain adaptation.

However, in basic CF models, there is no data with specific semantics (such as images and text), and we extract high-level dense features by embedding users and items into the latent space. In this way, we map users and items from different domains into different latent spaces. Taking movies as an example in Figure \ref{subfig:embedding_align}, solid and dotted lines indicate horror movies and comedies respectively. As shown in Figure \ref{subfig:embedding_align}, aligning embeddings directly may cause misleading --- horror movies from the ``orange'' domain and comedies from the ``blue'' domain are gathered and the distribution pattern are transferred incorrectly. The reason is that these embeddings are mapped into different latent spaces --- on the ``orange'' domain, the negative and positive semi-axes encode horrible and funny respectively, and on the ``blue'' domain, we face the opposite situation. To address this gap, we need to conduct domain adaptation in the same space, i.e., we align the spaces as well as embeddings.

To align the latent spaces, we explore domain-invariant features as anchor points. In this paper, we leverage textual features, which can be easily extracted from the user reviews. An example is shown in Figure \ref{subfig:embedding_text_align}. We concatenate the textual features with the embeddings thus the space is extended to a textual-latent space (the transverse axis indicates the latent space and longitudinal axis indicates the textual space). In the latent space shown in Figure \ref{subfig:embedding_align}, different categories are not separable, while in Figure \ref{subfig:embedding_text_align}, different categories are separable by extending textual dimensions. For domain adaptation, we use the concatenated embeddings and textual features as the input of the domain classifier. Embeddings are trained adversarially with the classifier while textual features are fixed.

As we can see, the textual features in our strategy should be domain-invariant, e.g., horror movies from all domains are mapped to the negative semi-axis of the textual space. There are many existing models extracting textual features for recommendation \cite{text_cnn,text_cnn_att}, while these features are not domain-invariant. To close this gap, we first propose a memory structure called \textbf{T}ext \textbf{M}emory \textbf{N}etwork (\textbf{TMN}) to extract textual features by mapping each user and item into the word semantic space. Then, we inject the features into a CF model to generate the prediction. The model consisting of the textual features and CF module is named as \textbf{T}ext \textbf{C}ollaborative \textbf{F}iltering (\textbf{TCF}) model. Finally, we train two TCF models on source and target domains synchronously, and connect them by an adaptation net. The transfer learning model is called \textbf{T}ext-enhanced \textbf{D}omain \textbf{A}daptation \textbf{R}ecommendation (\textbf{TDAR}) method.

Specifically, our contributions are listed as follows:

\begin{itemize}
	{\item We propose a domain adaptation recommendation method (TDAR) by aligning embeddings into a same latent space, which greatly improves the performances on the sparse datasets. To align the spaces and embeddings, we use the textual features as anchor points.}
	
	{\item As one important module in TDAR, we devise a memory network to extract domain-invariant textual features, and inject the features into a CF model to propose a text-based CF model.}
	
	{\item We devise comprehensive experiments on three pairs of real-world datasets to demonstrate the effectiveness of our proposed methods. Codes are available on \url{https://github.com/Wenhui-Yu/TDAR}.}
\end{itemize}

\section{Related Work}
\label{sec:related_work}
Recently, Recommender System (RS) has gained increasing attention due to its wide application on various online platforms. We model user preferences from the history interactions, and return personalized recommendation to each user. In various RS models, CF models mine collaborative information from the user-item interaction graph in a direct \cite{user-based,item-based} and a high-level \cite{MF,FM} way. Latent factor models \cite{MF,FM,NCF}, which are considered as a special kind of CF models, encode user preferences and item properties with embeddings, and measure the distance of embeddings in the latent space. To improve the representation capability, many variants have been proposed \cite{AES,text_cnn,WBPR,NGCF,LCFN}. Though extensively studied, there is still a critical issue: RS suffers from serious sparsity problem and the performance on sparse data leaves much to be desired. In this paper, we aim to leverage transfer learning to enrich the information on sparse data and we use textual features as the anchor points, thus we introduce the most relevant aspects: cross-domain recommendation, and text-based recommendation in this section.

\subsection{Cross-domain Recommendation}
\label{subsec:cross_domain_rec}
In this subsection, we first introduce the core technique --- domain adaptation, and then introduce some related work about cross-domain recommendation with and without overlap.

\subsubsection{Domain Adaptation} 
\label{subsubsec:domain_adaptation}
To learn on the data with fragmentary labels, or even without labels, transfer learning was proposed by transferring the knowledge learned from a well labeled source dataset to the target dataset \cite{domain_adaption1,MMD,selective_transfer,domain_adaption2}. \citet{domain_adaption2} proposed DANN for unsupervised image classification tasks. Assume $\{\bm{{{\rm x}}}^d_i, y^d_i\}_{i=1,\cdots, m; d = S}$ are labeled source data and $\{\bm{{{\rm x}}}^d_i\}_{i=1,\cdots, n; d = T}$ are unlabeled target data. There are three parts in DANN: a feature extractor $G_f(\;,\theta_f)$ (convolutional layers of CNN), a label predictor $G_y(\;,\theta_y)$ (fully-connected layers of CNN), and a domain classifier $G_d(\;,\theta_d)$. When training the model, $\theta_f$ and $\theta_y$ are updated to minimize the label prediction loss $\sum_{i}L(G_y(G_f(\bm{{{\rm x}}}^S_i,\theta_f),\theta_y), y^S_i)$ on the source domain, $\theta_d$ and $\theta_f$ are trained to minimize and maximize the domain prediction loss $\sum_{i,d}L(G_d(G_f(\bm{{{\rm x}}}_i^d,\theta_f),\theta_d),d)$ respectively. By adversarially training $\theta_d$ and $\theta_f$, visual features from the two domains $G_f(\bm{{{\rm x}}}_i^S,\theta_f)$ and $G_f(\bm{{{\rm x}}}_i^T,\theta_f)$ are aligned and the knowledge learned from the source domain is transferred to the target domain.

\subsubsection{Cross-domain Recommendation with Overlap} There are many models proposed for cross-domain recommendation with user and item overlap. \citet{transfer_rec} reduced the difference between source embeddings and target embeddings by directly minimizing the Frobenius norm of the difference. \citet{selective} proposed a selective transfer learning method which determines what to transfer based on a boost algorithm. \citet{CoNet} proposed two deep neural networks on source and target domains. By sharing user embedding layer, all users and items are mapped into the same latent space, and by constructing cross connections between the two networks, the parameters are transferred across domains. \citet{auto_encode_trans} encoded each user by an auto-encoder and aligned the user representations by domain adaptation. \citet{transfer_hybrid} used items on the source domain which have interactions with current user to enhance the representation of the current item on the target domain, and used user reviews to improve the model performance. We can see that transfer learning is easy to achieve since there is user and item overlap. In this case, the bipartite user-item graphs of the two domains are different parts of a whole graph, and we can transfer the knowledge by simply sharing embeddings of the overlapped users and items between the two domains. However, if there is no overlap, two graphs are totally separate and embeddings are not sharable, thus we have to utilize more advanced approaches such as domain adaptation.

\subsubsection{Cross-domain Recommendation without Overlap} There are several models for cross-domain recommendation without overlap. \citet{auto_encoder_text} introduced an interesting task --- recommending items from the source domain to users from the target domain based on text. To do so, textual features are extracted by an auto-encoder and aligned by domain adaptation. \citet{RSDAN} proposed Long Short-Term Memory (LSTM) to construct user and item textual representations, and aligned them of the two domains for transfer learning. As we can see, these models \cite{auto_encoder_text,RSDAN} achieve domain adaptation in the textual space since embeddings are difficult to align without overlap. In this case, cross-domain recommendation in \cite{auto_encoder_text,RSDAN} is more close to a Natural Language Processing (NLP) task rather than a recommendation task. \citet{codebook} proposed a ``codebook'' method which transfers the rating pattern in the cluster level, nevertheless too coarse and empirical. Moreover, this method is based on users' rating pattern thus is difficult to be extended to implicit feedback situations. In this paper, we aim to refine embeddings which are key representations of CF models. To the best of our knowledge, this paper is the first work focusing on embedding aligning for cross-domain recommendation tasks without user and item overlap.

\subsection{Text-based recommendation}
Since we want to use textual features as the anchor points to align embeddings, we extract domain-invariant textual features for each user and item. In this subsection, we introduce some text-based recommendation models. The basic models introduced in \cite{auto_encoder_text,RSDAN} are all text-based models. \citet{auto_encoder_text} devised an auto-encoder and \citet{auto_encoder_text} leveraged LSTM to extract user and item textual representations. \citet{text_cnn} gathered reviews for each user and item, and extract textual features from these reviews by a Convolutional Neural Network (CNN) and \citet{text_cnn_att} further added an attention mechanism. However, textual features extracted by these existing models are not domain-invariant. Inspired by \cite{transfer_hybrid}, we propose a memory network for textual features in this paper.

\section{Text Memory Network}
\label{sec:TMN}
In this paper, bold uppercase letters refer to matrices. Assuming there are $M$ users and $N$ items in total, we use matrix $\bm{{{\rm R}}} \in \mathbb{R}^{M \times N}$ to denote the interactions between users and items. $\bm{{{\rm R}}}_{ui} = 1$ if user $u$ voted items $i$ and $\bm{{{\rm R}}}_{ui} = 0$ otherwise. Our task is to give prediction (denoted as $\hat{\bm{{{\rm R}}}}$) of the missing values (0 in $\bm{{{\rm R}}}$).

In this section, we construct user- and item-specific textual representation from the reviews. Please note that the model devised in this section is for signal domain and we extract textual features for source and target domains separately. Taking user $u$ as an example, we construct a review set $R_u = \bigcup_{i=1}^Nr_{ui}$, where $r_{ui}$ is a set containing words of $u$'s review towards $i$. If $u$ did not interact with $i$, $r_{ui}=\varnothing$. Similarly, the review set of item $i$ is $R_i = \bigcup_{u=1}^Mr_{ui}$. We use $W$ to denote the set of words: $W = \bigcup_{u=1}^M\bigcup_{i=1}^Nr_{ui}$, and use $H$ to denote the total number of words: $H = |W|$. Since we want to extract domain-invariant textual features (i.e, features from all domains are in the same space), we map all users and items into the word semantic space by linearly combining word semantic vectors of the reviews. $\bm{{{\rm S}}} \in \mathbb{R}^{H\times K_1}$ is the word semantic matrix\footnote{\url{https://code.google.com/archive/p/word2vec}} pretrained on GoogleNews corpus by word2vec \cite{word2vec}, and $\bm{{{\rm S}}}_w$ indicates semantic features of the word $w$. We use $\bm{{{\rm E}}}\in \mathbb{R}^{M\times K_1}$ and $\bm{{{\rm F}}}\in \mathbb{R}^{N\times K_1}$ to denote textual features we construct for users and items, respectively. Taking user $u$ as an example, $\bm{{{\rm E}}}_u = \sum_{w\in R_u} a_{uw} \bm{{{\rm S}}}_w,$ where $a_{uw}$ is the weight of word $w$ based on $u$'s semantic preferences. We propose a Text Memory Network (TMN) to calculate weights for users $\{a_{uw}\}$ and for items $\{a_{iv}\}$ to construct textual features from word semantics. 

For a user $u$ which prefers horror movies, $u$ may prefer relevant words (such as ``horrible'', ``frightened'', ``terrifying''), and has no interest in irrelevant words (such as ``this'', ``is'', ``a'') and opposite words (such as ``funny'', ``relaxing'', ``comical''). For the word $w$ preferred by $u$, we need to set a large weight $a_{uw}$ for $w$. In the item aspect, for a horror movie $i$, the relevant words in $i$'s reviews provide much information about $i$ and the irrelevant or opposite words provide little information. For a word $v$ which is important to $i$, we need to set a large weight $a_{iv}$. It is easy to see that in this task, we aim to recommend preferred words to users and items. 

Inspired by matrix factorization \cite{MF}, we declare three matrices $\bm{{{\rm P}}}\in \mathbb{R}^{M\times K_2} $, $\bm{{{\rm Q}}}\in \mathbb{R}^{N\times K_2}$, and $\bm{{{\rm T}}}\in \mathbb{R}^{H\times K_2}$ for users, items, and words, respectively. Taking users as an example, we use $e_{uw}=\bm{{{\rm P}}}_u\bm{{{\rm T}}}_w^\mathsf{T}$ to model $u$'s preferences towards word $w$. To further emphasize important words, we input $\{e_{uw}\}$ to a softmax function to get $\{a_{uw}\}$: $a_{uw} = \frac{\exp(e_{uw})}{\sum_{w'\in R_u}\exp(e_{uw'})}$. Weights for items $\{a_{iv}\}$ are constructed in the same way. We finally predict user preferences towards items by $\hat{\bm{{{\rm R}}}} = \sigma(\bm{{{\rm E}}}\bm{{{\rm F}}}^\mathsf{T})$, where $\sigma(\;)$ is the element wise sigmoid function, and we use cross entropy loss as our loss function:
\begin{small}
	\begin{align}
	\label{equ:loss}
	\mathcal{L} = -\sum\limits_{u,i}\bm{{{\rm R}}}_{ui}\log\hat{\bm{{{\rm R}}}}_{ui} + (1-\bm{{{\rm R}}}_{ui})\log(1-\hat{\bm{{{\rm R}}}}_{ui})+\lambda reg,
	\end{align}
\end{small}where 
\begin{small}
	$$\hat{\bm{{{\rm R}}}}_{ui}=\sigma\Bigg[\Bigg(\sum\limits_{w\in R_u} \frac{\exp(\bm{{{\rm P}}}_u\bm{{{\rm T}}}_w^\mathsf{T})}{\sum\limits_{w'\in R_u}\exp(\bm{{{\rm P}}}_u\bm{{{\rm T}}}_{w'}^\mathsf{T})} \bm{{{\rm S}}}_w \Bigg) \Bigg(\sum\limits_{v\in R_i} \frac{\exp(\bm{{{\rm Q}}}_i\bm{{{\rm T}}}_v^\mathsf{T})}{\sum\limits_{v'\in R_i}\exp(\bm{{{\rm Q}}}_i\bm{{{\rm T}}}_{v'}^\mathsf{T})} \bm{{{\rm S}}}_v\Bigg)^\mathsf{T}\Bigg],$$
\end{small}and the regularization term $reg$ is the Frobenius norm of model parameters $\{\bm{{{\rm P}}},\bm{{{\rm Q}}},\bm{{{\rm T}}}\}$, which is learned by minimizing $\mathcal{L}$ with Adam \cite{adam}. In this text-based recommendation task, we get a tripartite graph of users, items, and words. When constructing the weights, we only leverage the connections of user-word and item-word. After constructing textual representations for users and items, we supervise the model by the user-item connections. As we can see, we indeed use three bipartite graphs in TMN.

Compared with existing CNN and RNN recommendation models \cite{text_cnn,RSDAN,text_cnn_att}, our textual feature extractor cannot model the sequential information while is good at highlighting important keywords. We consider that to extract interaction-specific textual information such as the task in \cite{Key_Frame}, sequential information is important. However, to extract user- and item-specific textual information such as the task in \cite{text_cnn,RSDAN,text_cnn_att} as well as in this paper, keywords are more important, since we want the textual features to summarize the preference elements of each user and item. Experiments also show that TMN performs very well in our task, especially in sparse cases.

\section{Text-enhanced Cross-domain Recommendation}
After extracting textual features by TMN, we introduce our Text-enhanced Domain Adaptation Recommendation (TDAR) model in this section. First, we inject the textual features into a CF model to propose the basic TCF model. Then, we train two TCF models (sharing the same interaction function, please see Figure \ref{fig:TDAR}) on target and source domains simultaneously, and align the user and item embeddings by domain adaptation.

\subsection{Text Collaborative Filtering}
\label{subsec:TCF}
In this subsection, we devise our basic model. $\bm{{{\rm U}}} \in \mathbb{R}^{M\times K_3}$ and $\bm{{{\rm V}}} \in \mathbb{R}^{N\times K_3}$ are embeddings for users and items respectively. As illustrated in Figure \ref{subfig:embedding_text_align}, we concatenate embeddings and textual features thus the representations of users and items are $[\bm{{{\rm U}}}, \bm{{{\rm E}}}]$ and $[\bm{{{\rm V}}}, \bm{{{\rm F}}}]$. We predict user preferences by these representations: 
\begin{small}
	$$\hat{\bm{{{\rm R}}}}_{ui}=f([\bm{{{\rm U}}}, \bm{{{\rm E}}}]_u,[\bm{{{\rm V}}}, \bm{{{\rm F}}}]_i, \bm{{{\rm \Theta}}}),$$
\end{small}where $f(\;, \bm{{{\rm \Theta}}})$ is the interaction function combining user and item embeddings and returning the preference prediction, such as a deep structure \cite{NCF,NGCF,LCFN}, and $\bm{{{\rm \Theta}}}$ indicates the parameters. We also learn the model by minimizing the loss function given in Equation (\ref{equ:loss}). In this model, $\bm{{{\rm U}}}$, $\bm{{{\rm V}}}$, and $\bm{{{\rm \Theta}}}$ are trainable parameters while $\bm{{{\rm E}}}$ and $\bm{{{\rm F}}}$ are fixed. $reg$ in Equation (\ref{equ:loss}) indicates the Frobenius norms of $\bm{{{\rm U}}}$ and $\bm{{{\rm V}}}$.

\subsection{Text-enhanced Domain Adaptation Recommendation}
\label{subsec:TDAR}
In this subsection, we use superscripts $s$, $t$, $u$ and $i$ to indicate source domain, target domain, user, and item, respectively. $\bm{{{\rm R}}}^s\in \mathbb{R}^{M^s\times N^s}$ and $\bm{{{\rm R}}}^t\in \mathbb{R}^{M^t\times N^t}$ indicate interactions on source and target domains. We train two TCFs on the two domains while share the same interaction function, thus the prediction on the two datasets is given by:
\begin{small}
	\begin{flalign}
	\label{equ:pre1}
	\hat{\bm{{{\rm R}}}}^s_{u^si^s}=f([\bm{{{\rm U}}}^s, \bm{{{\rm E}}}^s]_{u^s},[\bm{{{\rm V}}}^s, \bm{{{\rm F}}}^s]_{i^s}, \bm{{{\rm \Theta}}}),
	\end{flalign}
	\vspace{-4mm}
	\begin{flalign}
	\label{equ:pre2}
	\hat{\bm{{{\rm R}}}}^t_{u^ti^t}=f([\bm{{{\rm U}}}^t, \bm{{{\rm E}}}^t]_{u^t},[\bm{{{\rm V}}}^t, \bm{{{\rm F}}}^t]_{i^t}, \bm{{{\rm \Theta}}}).
	\end{flalign}
\end{small}We then add adaptation networks on the two TCFs to achieve transfer learning. Considering the distribution pattern of user and item embeddings are possibly different, we conduct domain adaptation for users and items separately. Here we take users as an example. Assuming there are two user embedding distributions $dist(\bm{{{\rm U}}}^s)$ and $dist(\bm{{{\rm U}}}^t)$, we use a binary variable $d^u_u$ as the domain label, which indicates whether $\bm{{{\rm U}}}_{u}$ come from the target distribution or from the source distribution: $d^u_u = 1$ if $\bm{{{\rm U}}}_{u}\sim dist(\bm{{{\rm U}}}^t)$ and $d^u_u = 0$ if $\bm{{{\rm U}}}_{u}\sim dist(\bm{{{\rm U}}}^s)$. The superscript of $d^u_u$ indicates that the domain label is for user embeddings and the subscript indicates that the domain label is for the current user $u$.

For the adaptation net, we leverage a domain classifier denoted as $g(\;,\bm{{{\rm \Phi}}}^u)$, which is trained for domain classification: $\hat{d}^u_u = g([\bm{{{\rm U}}},\bm{{{\rm E}}}]_u,$ $\bm{{{\rm \Phi}}}^u)$. To align embeddings, we want the distributions $dist(\bm{{{\rm U}}}^s)$ and $dist(\bm{{{\rm U}}}^t)$ to be similar. The most widely-used way is to train the domain classifier to discriminate between the two distributions, and train embeddings to puzzle the classifier \cite{domain_adaption1,domain_adaption2}. To be specific, we update $\bm{{{\rm \Phi}}}^u$ to minimize the loss of $g(\;,\bm{{{\rm \Phi}}}^u)$ and update $\bm{{{\rm U}}}^s$ and $\bm{{{\rm U}}}^t$ to maximize it. In this way, the user embeddings of two domains are not separable therefore are aligned to the same distribution. Item embeddings are aligned in the same way.

\begin{figure}[ht]
	\centering
	\includegraphics[scale = 0.52]{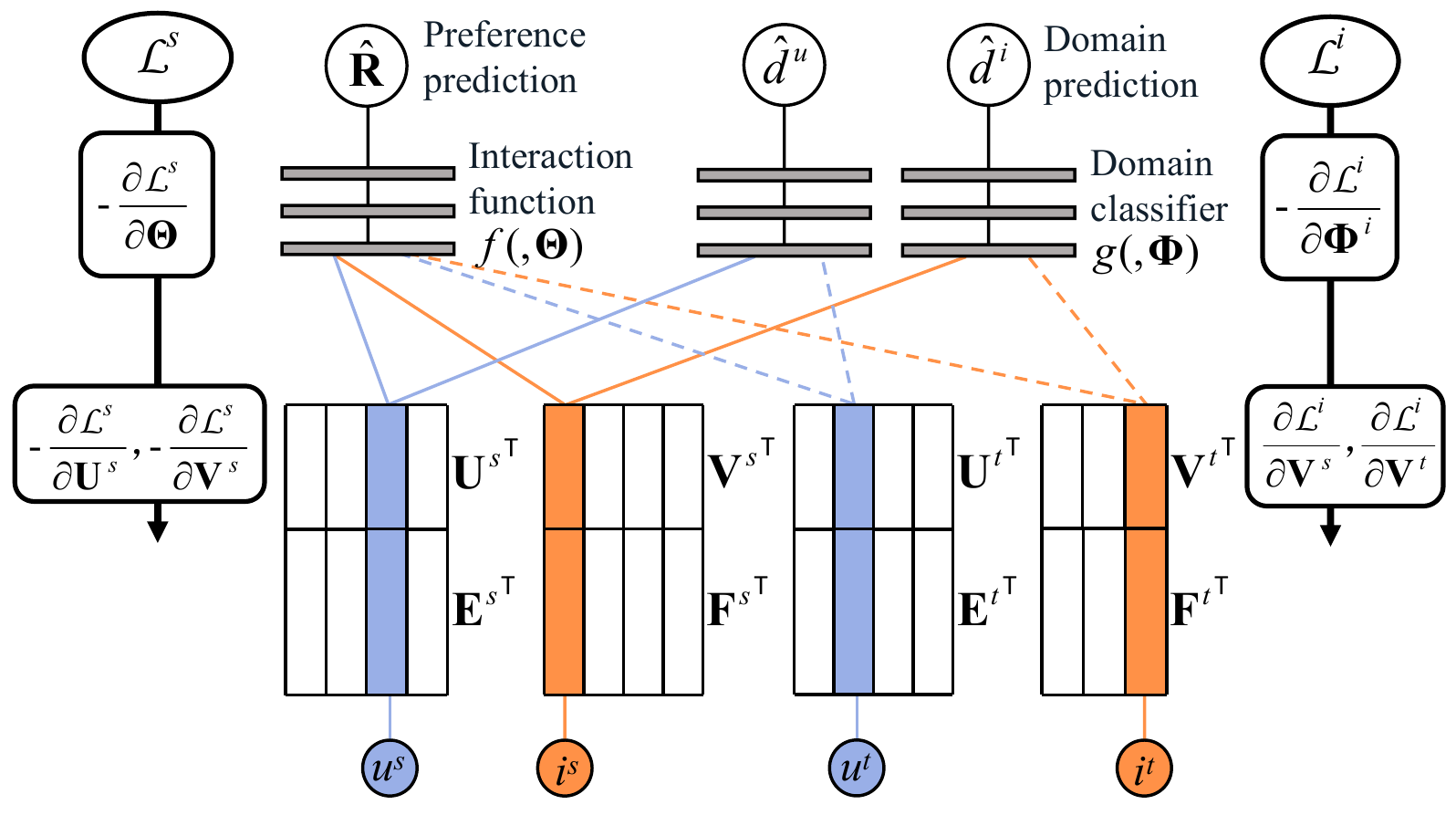}
	\vspace{-2mm}
	\caption{Illustration of TDAR. $\bm{{{\rm U}}}$ and $\bm{{{\rm V}}}$ are user and item embeddings and $\bm{{{\rm E}}}$ and $\bm{{{\rm F}}}$ are user and item textual features. $f(\;, \bm{{{\rm \Theta}}})$ is the interaction function and $g(\;, \bm{{{\rm \Phi}}})$ is the domain classifier. Lines indicate the forward propagation of the model and we use the solid lines and dotted lines to distinguish between different samples. Thick arrows indicate the backward propagation. We only show the prediction loss on the source domain $\mathcal{L}^s$ and classification loss for items $\mathcal{L}^i$ and corresponding gradients for concise illustration. \vspace{-3mm}}
	\label{fig:TDAR}
\end{figure}

We use $\mathcal{L}^s$ and $\mathcal{L}^t$ to denote prediction loss on the source domain and target domain, and use $\mathcal{L}^u$ and $\mathcal{L}^i$ to denote domain classification loss for users and items, respectively. $\mathcal{L}^s$, $\mathcal{L}^u$, and $\mathcal{L}^i$ are all cross entropy loss of the binary predictors. For $\mathcal{L}^t$, we only use positive labels as supervision on the target domain. The loss functions are listed as follows:
\begin{small}
	\begin{flalign}
	\label{equ:losses}
	\left\{
	\begin{array}{l}
	\mathcal{L}^s = -\!\sum\limits_{u^s,i^s}\!\bm{{{\rm R}}}^s_{u^si^s}\!\log\!\hat{\bm{{{\rm R}}}}^s_{u^si^s} \!+\! (1\!-\!\bm{{{\rm R}}}^s_{u^si^s})\!\log(1\!-\!\hat{\bm{{{\rm R}}}}^s_{u^si^s})\! +\! \lambda^s reg^s\\
	\mathcal{L}^t = -\sum\limits_{u^t,i^t}\bm{{{\rm R}}}^t_{u^ti^t}\log\hat{\bm{{{\rm R}}}}^t_{u^ti^t}+ \lambda^t reg^t\\
	\mathcal{L}^u = -\sum\limits_{u^s,u^t}\log\hat{d}^u_{u^t} + \log(1-\hat{d}^u_{u^s})\\
	\mathcal{L}^i = -\sum\limits_{i^s,i^t}\log\hat{d}^i_{i^t} + \log(1-\hat{d}^i_{i^s})\\
	\end{array},
	\right.
	\end{flalign}
\end{small}where, $u^s$ and $i^s$ are the user and item from the source domain, and $u^t$ and $i^t$ are that from the target domain. $\hat{\bm{{{\rm R}}}}^s_{u^si^s}$ and $\hat{\bm{{{\rm R}}}}^t_{u^ti^t}$ are given in Equations (\ref{equ:pre1}) and (\ref{equ:pre2}). $\hat{d}^u_u$ and $\hat{d}^i_i$ are given by $\hat{d}^u_u = g([\bm{{{\rm U}}},\bm{{{\rm E}}}]_u,$ $\bm{{{\rm \Phi}}}^u)$ and $\hat{d}^i_i = g([\bm{{{\rm V}}},\bm{{{\rm F}}}]_i,$ $\bm{{{\rm \Phi}}}^i)$. Please note that the classifiers $g(\;, \bm{{{\rm \Phi}}}^u)$ and $g(\;, \bm{{{\rm \Phi}}}^i)$ share the same structure while with different parameters. $reg^s$ and $reg^t$ indicate the Frobenius norms of $\{\bm{{{\rm U}}}^s,\bm{{{\rm V}}}^s\}$ and $\{\bm{{{\rm U}}}^t,\bm{{{\rm V}}}^t\}$ respectively, and $\lambda^s$ and $\lambda^t$ are corresponding regularization coefficients. We update the model parameters as follows:
\begin{small}
	\begin{flalign}
	\label{equ:grad1}
	{\bm{{{\rm \Theta}}}}\leftarrow \bm{{{\rm \Theta}}}-\nabla_{\bm{{{\rm \Theta}}}} (\eta^s\mathcal{L}^s + \eta^t \mathcal{L}^t),
	\end{flalign}
	\vspace{-4mm}
	\begin{flalign}
	\label{equ:grad2}
	\bm{{{\rm U}}}^s/\bm{{{\rm V}}}^s\leftarrow \bm{{{\rm U}}}^s-\nabla_{{\bm{{{\rm U}}}}^s} (\eta^s \mathcal{L}^s - \eta^- \mathcal{L}^u) / \bm{{{\rm V}}}^s-\nabla_{{\bm{{{\rm V}}}}^s} (\eta^s \mathcal{L}^s - \eta^- \mathcal{L}^i),
	\end{flalign}
	\vspace{-4mm}
	\begin{flalign}
	\label{equ:grad3}
	\bm{{{\rm U}}}^t/\bm{{{\rm V}}}^t\leftarrow \bm{{{\rm U}}}^t-\nabla_{\bm{{{\rm U}}}^t}(\eta^t \mathcal{L}^t -\eta^- \mathcal{L}^u) / \bm{{{\rm V}}}^t-\nabla_{\bm{{{\rm V}}}^t} (\eta^t \mathcal{L}^t-\eta^-\mathcal{L}^i),
	\end{flalign}
	\vspace{-4mm}
	\begin{flalign}
	\label{equ:grad4}
	\bm{{{\rm \Phi}}}^u/\bm{{{\rm \Phi}}}^i\leftarrow \bm{{{\rm \Phi}}}^u-\eta^+\nabla_{\bm{{{\rm \Phi}}}^u} \mathcal{L}^u / \bm{{{\rm \Phi}}}^i-\eta^+ \nabla_{\bm{{{\rm \Phi}}}^i} \mathcal{L}^i,
	\end{flalign}
\end{small}where $\eta^s$, $\eta^t$, $\eta^+$ and $\eta^-$ are learning rates, and $\nabla_{\bm{{{\rm X}}}}f(\bm{{{\rm X}}})$ is the gradient of $f(\bm{{{\rm X}}})$ with respect to $\bm{{{\rm X}}}$. All parameters in TDAR are trained with Adam as well. The structure of TDAR is shown in Figure \ref{fig:TDAR}.

In TDAR, the two basic models are supervised by both labels and the domain classifier. As discussed in Introduction, negative labels on the target domain are contaminated by label noise seriously nevertheless the positive labels are pure, hence we abandon negative samples. Unfortunately, supervision with only positive samples in implicit feedback cases leads to a new problem --- the model tends to predict all items as positive items. To deal with this issue, we leverage the domain adaptation mechanism to supervise the basic model on the target domain together with positive samples. However, as a dual-domain system, negative sampling for the whole system is necessary. Considering that the source domain is usually much denser than the target domain and the quality of negative labels is much higher, we adopt the source domain for negative supervision, and transfer the negative supervision to the target domain by domain adaptation. 

For the textual features $\bm{{{\rm E}}}$ and $\bm{{{\rm F}}}$, we can pretrain and fix them during TDAR training and can also train them jointly with TDAR from scratch. Experiments show that joint training makes the model more difficult to tune yet achieves no performance improvement, so we choose the former strategy. Experiments also show that learning textual features is much more robust to label noise than learning user and item embeddings (please see Figure \ref{fig:experiment_result}, especially obvious in Figures \ref{subfig:cd_f1} and \ref{subfig:clothes_f1}). As a result, for the textual feature extractor --- TMN, we train it dependently on the two domains without aforementioned transfer learning strategy. The motivation is that we aim to use the textual features to guide embedding aligning hence do not want to align them blindly. For example, textual representations of two different domains (such as movies and clothes) should be different.

\section{Experiment}
In this section, we conduct experiments to validate the effectiveness of our model. We utilize three pairs of real-world datasets to compare our models against several state-of-the-art models, and report the experiment result. We focus on answering three following research questions:

\partitle{RQ1} How is the performance of our TDAR model?

\partitle{RQ2} How is the performance of our TMN and TCF models?

\partitle{RQ3} How is the effectiveness of our text-enhanced domain adaptation strategy?

\subsection{Experimental Setup}
In this subsection, we introduce some details about experimental setup.

\begin{table}[ht]
	\caption{Statistics of datasets}
	\centering
	\label{tab:datasets}
	\scalebox{0.9}{
		\begin{tabular}{cccccc}
			\toprule[1.2pt]
			Dataset & Interaction & User & Item & Sparsity & Source/Target \\
			\hline
			\textit{Movies} & 1,697,532 & 123,960 & 50,052 & 99.9726\% & Source \\
			\textit{Videos} & 11,137 & 4,555 & 1,580 & 99.8453\% & Target \\
			\textit{CDs} & 43,903 & 25,400 & 24,904 & 99.9931\% & Target \\
			\textit{Clothes} & 89,176 & 35,669 & 21,554 & 99.9884\% & Target \\
			\bottomrule[1.2pt]
	\end{tabular}}
\end{table}

\subsubsection{Datasets} 
\label{subsubsec:dataset}
We adopt \textit{Amazon}\footnote{\url{http://snap.stanford.edu/data/amazon/productGraph/}} dataset, which is the user reviews collected from the E-commerce website \textit{Amazon.com}, to evaluate our model. We select four categories of the \textit{Amazon} dataset: \textit{Movies}, \textit{Videos}, \textit{CDs}, and \textit{Clothes}. Statistics of these datasets are illustrated in Table \ref{tab:datasets}. All datasets are filtered by 5-core (removing items and users less than 5 records). We select the largest one --- \textit{Movies} as the source domain and the others as target domains. Since we want a dense source dataset, we filter \textit{Movies} by 30-core to increase the density. For target datasets, we want them to simulate real-world applications which contain many cold users and items (with less than 5 records). To close this gap, we randomly delete a part of interactions in the three target datasets. Since \textit{Amazon} is explicit feedback data (ratings), we set the interaction $(u,i)$ as ``1'' if $u$ has rated $i$ and ``0'' otherwise to construct implicit feedbacks. Each dataset is split into training set (80\%), validation set (10\%), and test set (10\%) randomly. We train models on the training set; determine all hyperparameters on the validation set; and report the performance on the test set. The target datasets are selected on consideration of different sparsity and similarity to the source dataset. To our intuition, \textit{Videos}, \textit{CDs}, and \textit{Clothes} are very similar, relatively similar, and different to \textit{Movies}, respectively.

\subsubsection{Model Setting}
$g(\;)$ is a deep fully-connected (FC) structure with 4 layers. The size of input layer is $K_1+K_3$ and of FC layers is 64. For the interaction function $f(\;, \bm{{{\rm \Theta}}})$, we adopt the inner product activated by sigmoid, i.e., $f(\bm{{{\rm U}}}_u, \bm{{{\rm V}}}_i, \bm{{{\rm \Theta}}}) = \sigma(\bm{{{\rm U}}}_u \bm{{{\rm V}}}_i^\mathsf{T})$, where $\bm{{{\rm \Theta}}}=\varnothing$. We choose this shallow and simple structure since we gain no performance enhancement by deep recommendation models in our experiments (please see Figure \ref{fig:experiment_result} for details).

\subsubsection{Baselines}
\label{subsubsec:baselines}
We adopt the following methods as baselines for comparison:

\begin{itemize}
	\item{\textbf{ItemPop:} This method ranks items based on \textbf{Item} \textbf{Pop}u-larity. It is a non-personalized method to benchmark the recommendation performance \cite{BPR,NCF}.}
	\item{\textbf{MF:} This \textbf{M}atrix \textbf{F}actorization \cite{MF} method is the basic yet competitive model. MF uncovers the underlying latent factors that encode the user preferences and item properties to predict missing values.}
	\item{\textbf{NeuMF:} This \textbf{Neu}ral \textbf{M}atrix \textbf{F}actorization method is the state-of-the-art neural network method for recommendation \cite{NCF}. NeuMF learns non-linear combination of user and item embeddings by wide and deep structures.}
	\item{\textbf{CoNN:} This Deep \textbf{Co}operative \textbf{N}eural \textbf{N}etworks is the state-of-the-art text-based recommendation method \cite{text_cnn}. CoNN leveraged CNN to extract user and item textual features from reviews to predict the user preferences.}
	\item{\textbf{DANN:} This \textbf{D}omain \textbf{A}dversarial \textbf{N}eural \textbf{N}etwork (\textbf{DANN}) \cite{domain_adaption2} is proposed to align high-level representations in various machine learning tasks. We utilize it to align embeddings directly in the recommendation context. We use MF as the basic models, which are supervised on two domains.}
	\item{\textbf{Rec-DAN:} This \textbf{D}iscriminative \textbf{A}dversarial \textbf{N}etworks for \textbf{Rec}ommender Systems is the state-of-the-art cross-domain recommendation model \cite{RSDAN}. Rec-DAN used DANN to align textual features to transfer the knowledge. The target part of Rec-DAN is unsupervised.}
\end{itemize}

Since several baselines are designed for explicit feedbacks (MF, CoNN, Rec-DAN), we use cross entropy to optimize them in our experiments. Note that though we introduced many cross-domain recommendation models in Related Work, most of them are not comparable. Only \cite{RSDAN,codebook} are designed for the no overlap case. The codebook approach \cite{codebook} is widely used yet cannot be extend to implicit feedbacks, we leave out the comparison with it.

\subsubsection{Evaluation Protocols} To evaluate the performance of our proposed model and baselines in implicit feedback case, we rank all items for each user in validation/test set and recommend the top-$k$ items to the user. We then adopt two metrics, $F_1$-score and normalized discounted cumulative gain (NDCG) to evaluate the recommendation quality. $F_1$-score, which is defined as harmonic mean of precision and recall, is extensively used to test the accuracy of binary classifier. NDCG is a position-sensitive metric widely used to measure the ranking quality. We recommend top-$k$ items to each user and calculate metrics, and finally use the average metrics of all users to remark the performance of the models. 

\subsubsection{Parameter Setting}\label{subsubsec:para_setting} To compare fairly, all models in our experiments are tuned with the following strategies: The maximum iteration number is set to 200. In each iteration, we enumerate all positive samples and test the model. The learning rate $\eta$ and the regularization coefficient $\lambda$ are determined by grid search in the coarse grain range of $\{0.0001,0.001,0.01,0.1\}\otimes \{0.001,0.01, 0.1,1\}$ and then in the fine grain range, which is based on the result of coarse tuning. For example, if a certain model achieves the best performance when $\eta=0.01$ and $\lambda=0.1$, we then tune it in the range of $\{0.002, 0.005, 0.01, 0.02, 0.05\} \otimes \{0.02, 0.05, 0.1, 0.2, 0.5\}$. Considering some baselines (CoNN and Rec-DAN) use dropout for generalization, we tune these models with respect to $\eta$ and the dropout rate also by the coarse and fine grid search. We determine the batch size in range of $\{64, 128, \cdots, 1024\}$ and the embedding length in the range of $\{8,16,\cdots,128\}$. The experiments are repeated 5 times for each parameter setting in model tuning and 10 times for the final comparison. Our experiments are conducted by predicting Top-2, 5, 10, 20, 50, 100 favourite items and we tune models according to $F_1$-score@2.

The aforementioned part is the tuning strategy for single-domain models, now we take TDAR as example to introduce the tuning strategy of cross-domain models. We train and tune TCF on the source domain, and initialize source domain embeddings of TDAR ($\bm{{{\rm U}}}^s$ and $\bm{{{\rm V}}}^s$) by pretrained TCF. Also, hyperparameters ($\eta^s$ in Equations (\ref{equ:grad1}) and (\ref{equ:grad2}) and $\lambda^s$ in Equation (\ref{equ:losses})) are set as the best hyperparameter setting of pretrained TCF. For learning rates $\eta^+$, we set $\eta^s$, $\eta^t$, and $\eta^-$ as 0 in TDAR to train a embedding classification task to determine the best $\eta^+$. To determine $\eta^-$, we set $\eta^+$ as the best $\eta^+$ we got in last step and set $\eta^s$ and $\eta^t$ as 0 to train a embedding aligning task. $\eta^+$ and $\eta^-$ are both tuned in range of $\{1,0.1,\cdots,0.00001\}$ according to the classification accuracy of $g(\;)$. For remained hyperparameters $\eta^t$ and $\lambda^t$, we tune TDAR with respect to them by the tuning strategy designed for single-domain models.

\subsection{Performance of TDAR (RQ1)} 
\label{subsec:RQ1}

\begin{figure*}[ht]
	\centering
	\subfigure[\textit{Movies} --- $F_1$-score]{
		\includegraphics[scale = 0.22]{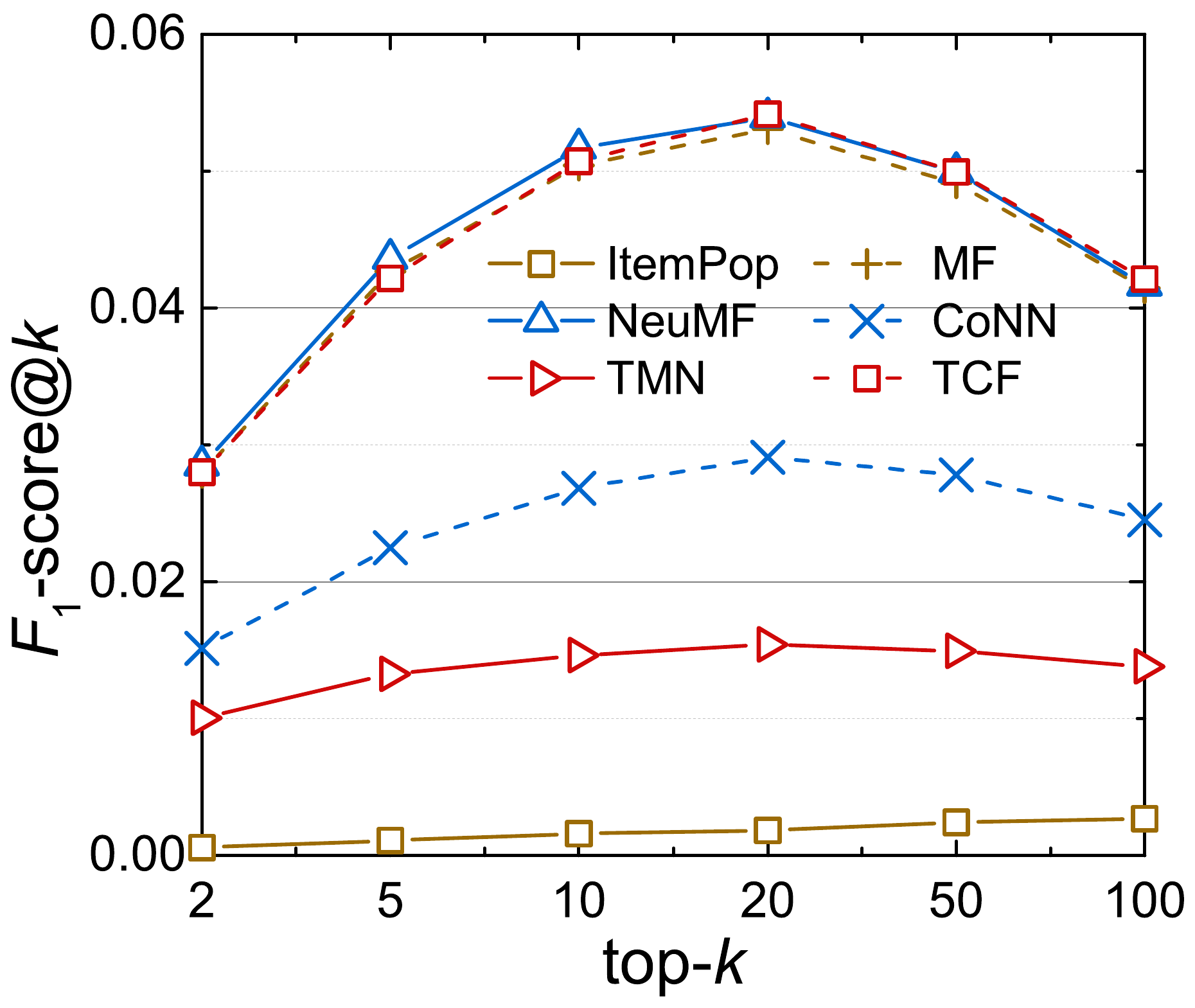}
		\label{subfig:movie_f1}
	}
	\subfigure[\textit{Videos} --- $F_1$-score]{
		\includegraphics[scale = 0.22]{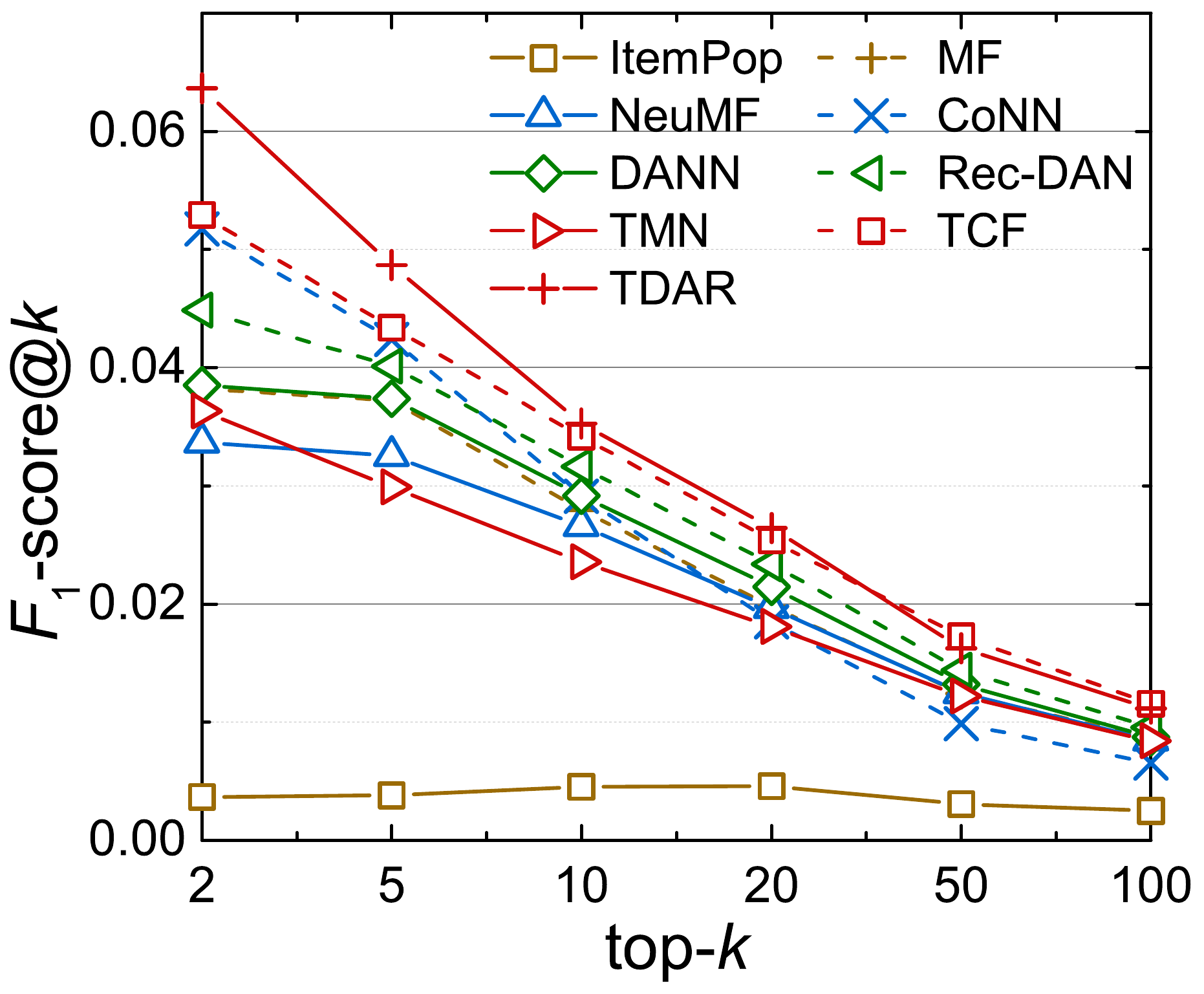}
		\label{subfig:video_f1}
	}
	\subfigure[\textit{CDs} --- $F_1$-score]{
		\includegraphics[scale = 0.22]{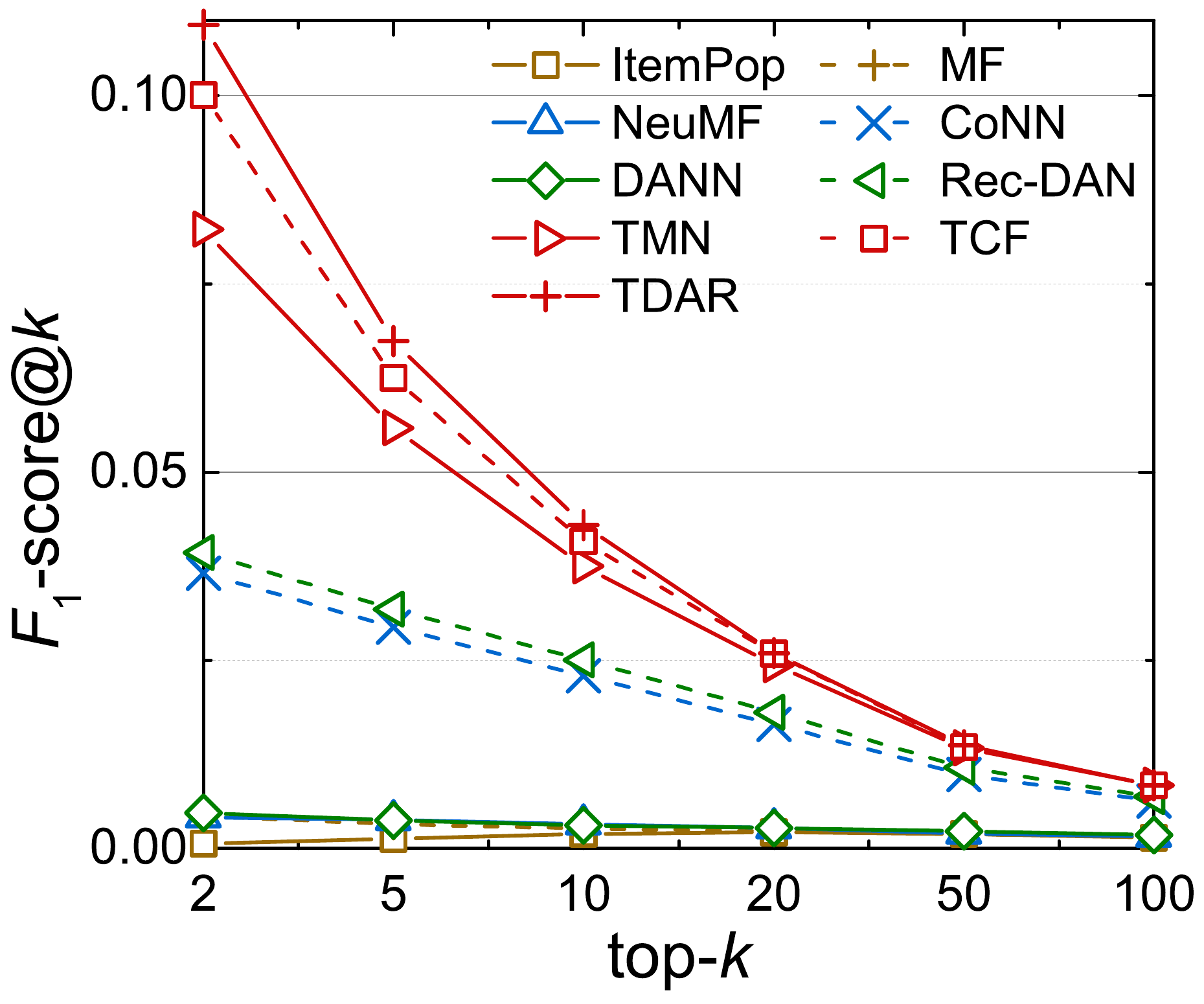}
		\label{subfig:cd_f1}
	}
	\subfigure[\textit{Clothes} --- $F_1$-score]{
		\includegraphics[scale = 0.22]{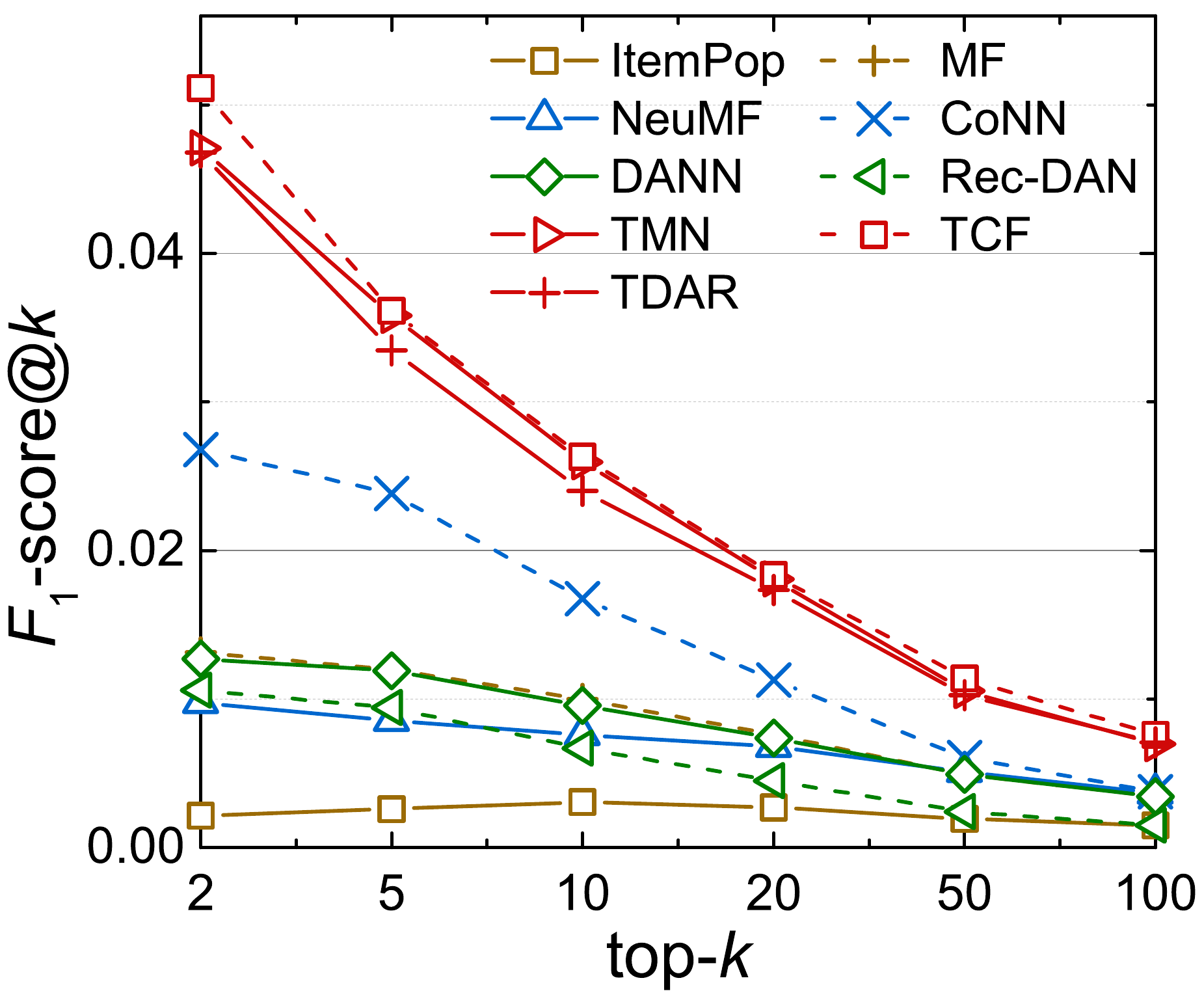}
		\label{subfig:clothes_f1}
	}
	\subfigure[\textit{Movies} --- NDCG]{
		\includegraphics[scale = 0.22]{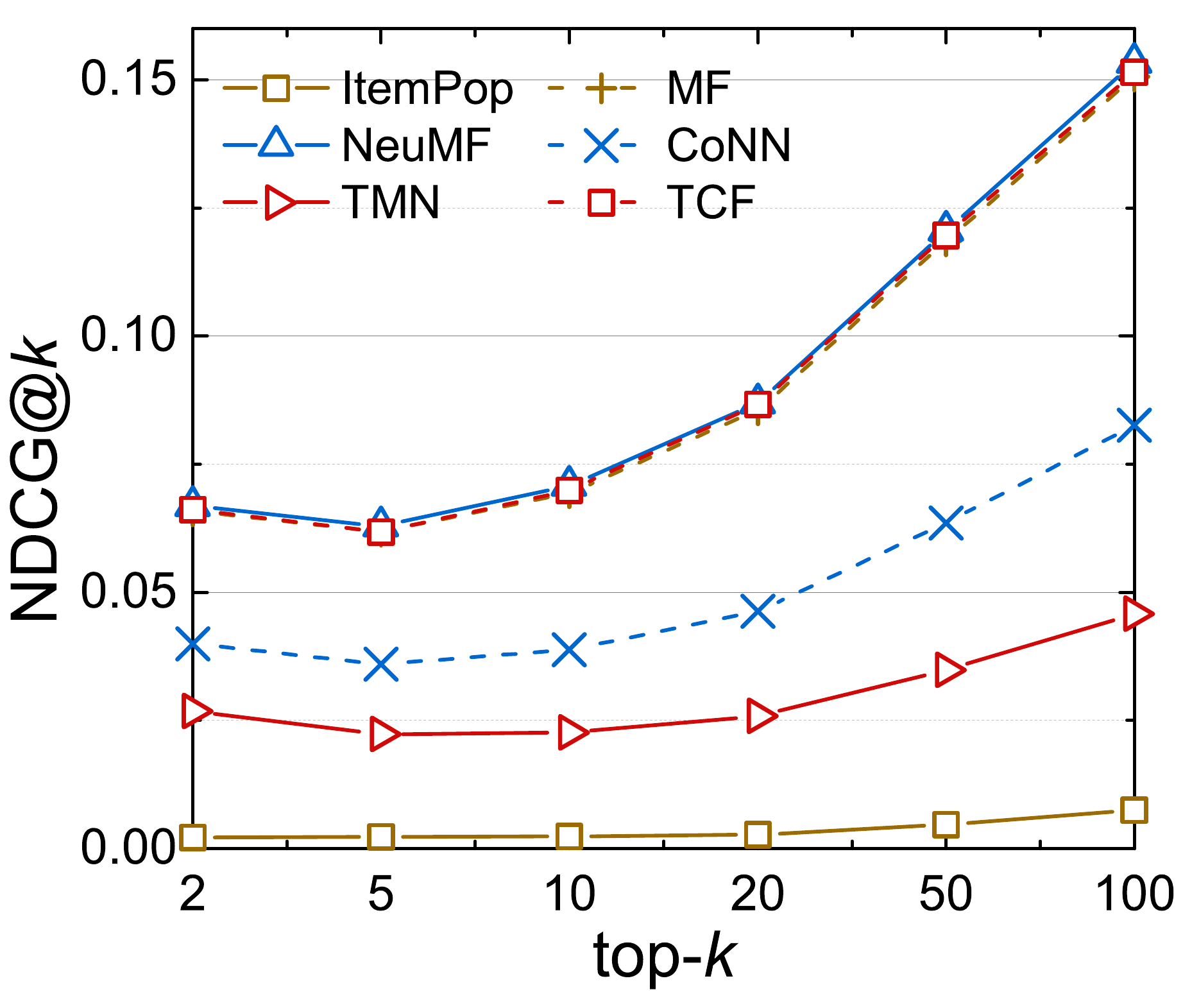}
		\label{subfig:movie_ndcg}
	}
	\subfigure[\textit{Videos} --- NDCG]{
		\includegraphics[scale = 0.22]{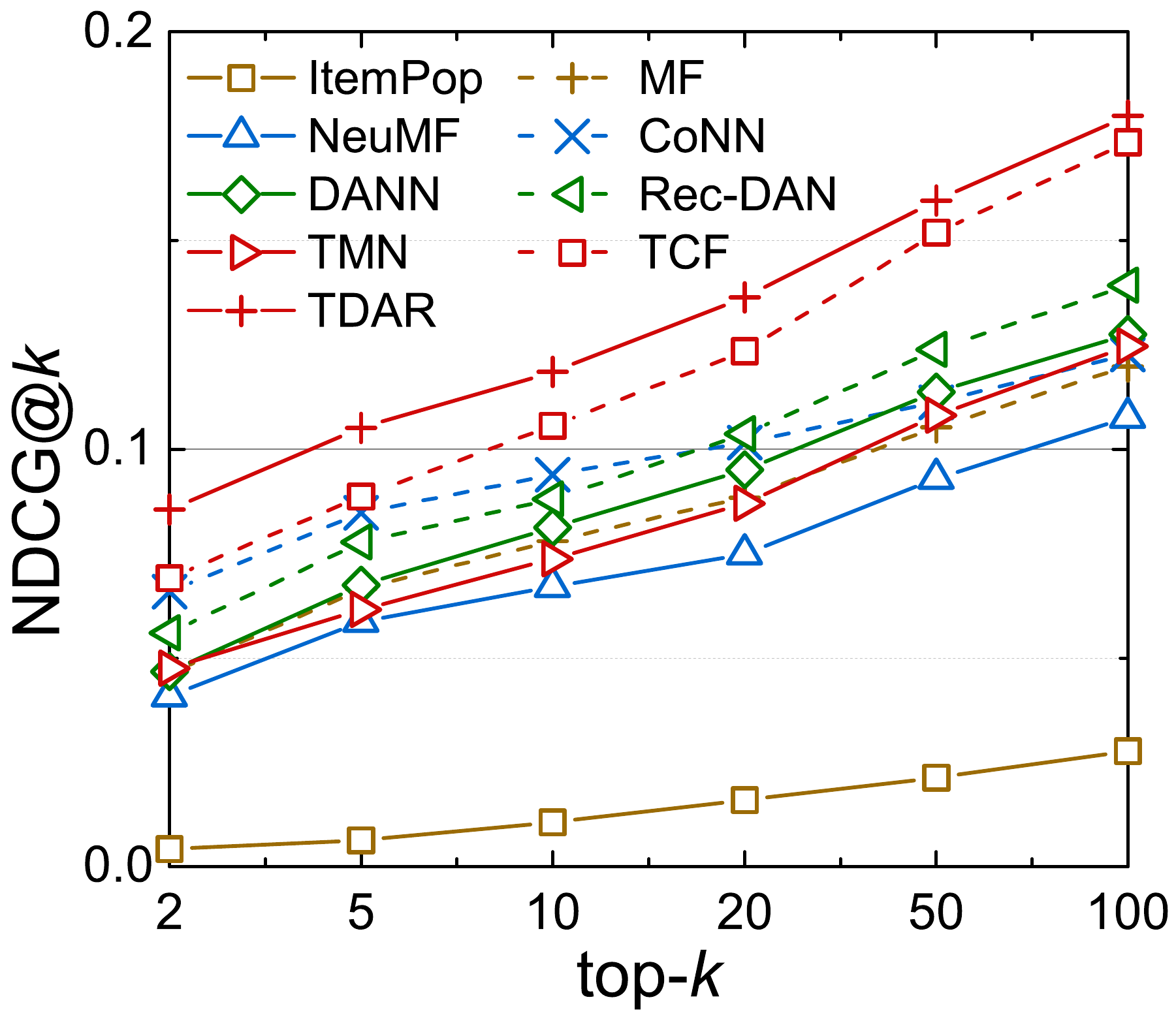}
		\label{subfig:video_ndcg}
	}
	\subfigure[\textit{CDs} --- NDCG]{
		\includegraphics[scale = 0.22]{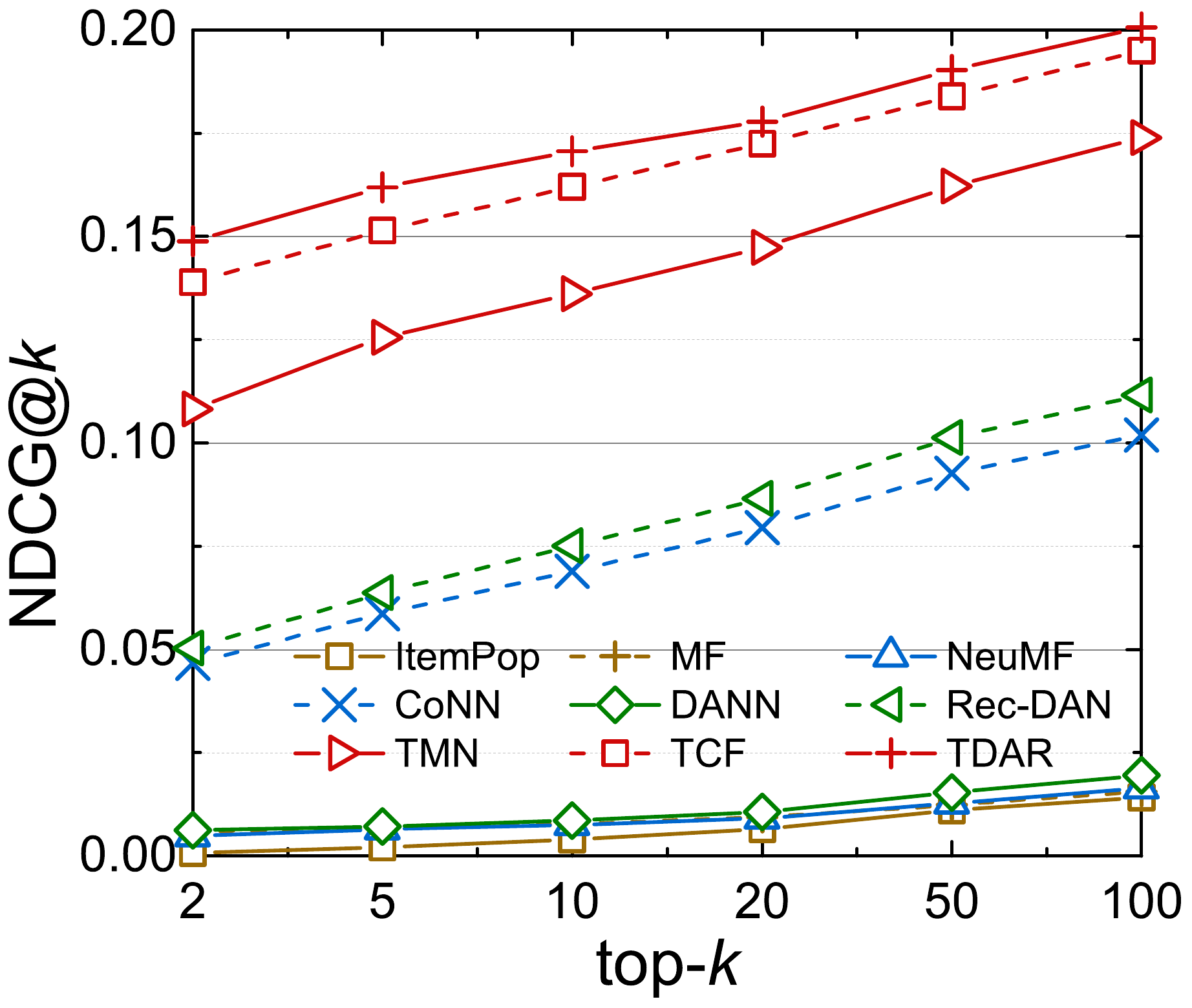}
		\label{subfig:cd_ndcg}
	}
	\subfigure[\textit{Clothes} --- NDCG]{
		\includegraphics[scale = 0.22]{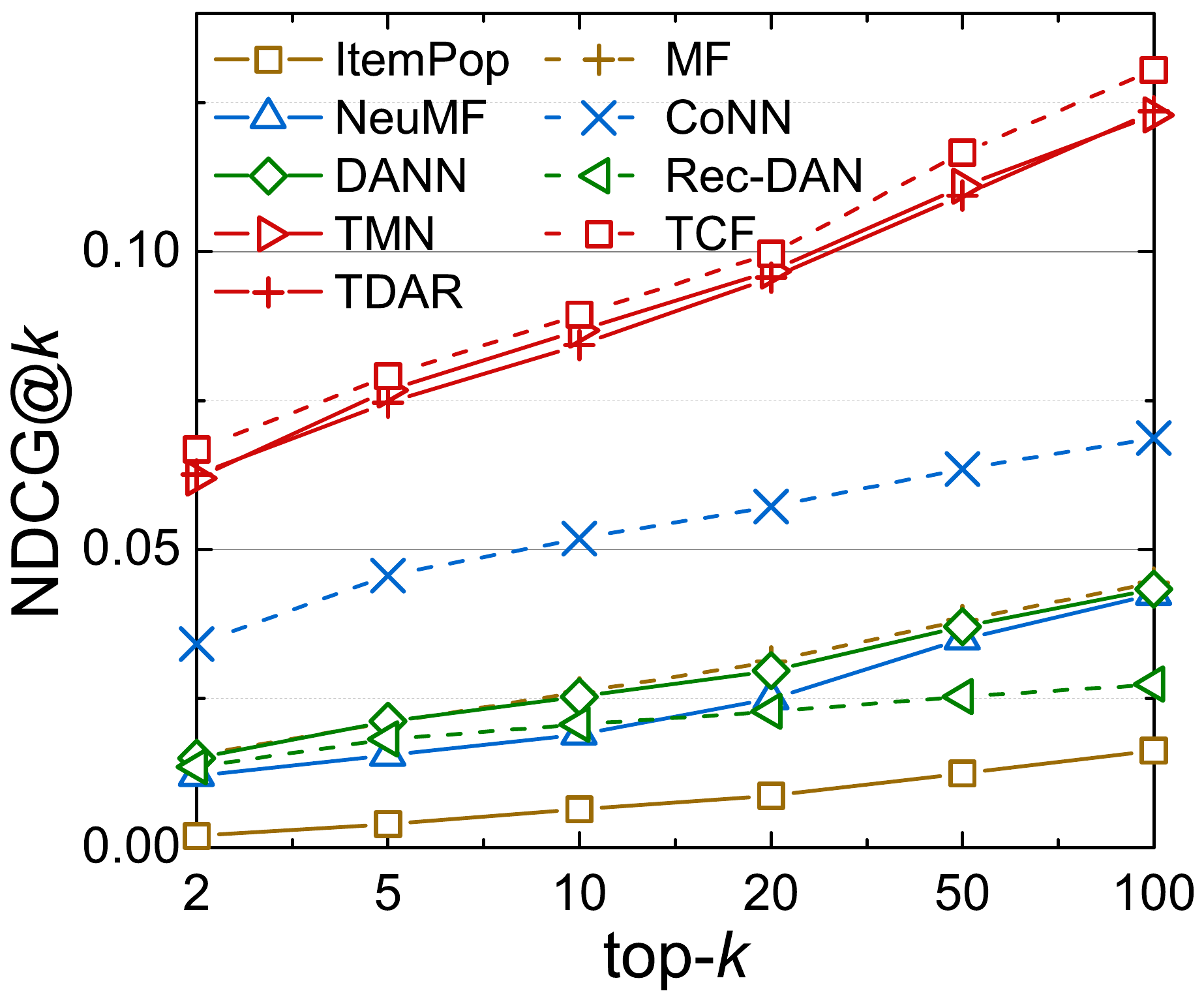}
		\label{subfig:clothes_ndcg}
	}
	\caption{Overall performance comparison (test set)}
	\label{fig:experiment_result}
\end{figure*}

We compare our models, TMN, TCF, and TDAR against several baselines and show the performance in Figure \ref{fig:experiment_result}. Lines in red, green, blue, and brown indicate our proposed models, cross-domain baselines, deep baselines, and shallow baselines, respectively. ItemPop is a non-personalized baseline to benchmark the recommendation performance on a dataset, and from its performance we can see that these three target datasets are very challenging due to the sparsity. Compared with ItemPop, all learning models capture users' personal preferences and outperform it significantly --- 8.23 times higher on $F_1$-score for the best case. MF is the basic learning baseline and NeuMF is the corresponding deep version, in which FC layers are leveraged to learn the interaction of user and item latent factors. NeuMF is a competitive model widely used in various recommendation scenarios and usually outperforms MF significantly \cite{NCF,NGCF}, however in our experiments, NeuMF fails to outperform MF. The possible reason is that the datasets we adopt are extremely sparse and NeuMF faces a more serious overfitting problem due to its stronger representation ability. Considering that our model is mainly designed for sparse datasets, we use inner product as our interaction function. 

CoNN is also a deep model for recommendation, which extracts textual features from user reviews. In our experiments, this text-based model outperforms latent factor models (MF and NeuMF) in most cases, thus we come to the conclusion that without sufficient supervision, side information boosts the performance significantly, because that latent factor models only get information from interactions while CoNN additionally gets information from user reviews. Comparing Table \ref{tab:datasets} and Figure \ref{fig:experiment_result}, we can see that on sparse datasets (especially on \textit{CDs}), interactions cannot provide enough information, thus the performance of latent factor models is extremely dreadful. Supported by textual information, CoNN shows more robust performance.

DANN and Rec-DAN are two cross-domain recommendation models by domain adaptation. DANN aligns latent factors and Rec-DAN aligns textual feature to transfer knowledge. As we discussed, latent factor models is uncompetitive compared with text-based models on sparse datasets, thus Rec-DAN performs much better than DANN --- 7.45 times improvement on $F_1$-score for the best case. Enhanced by both the novel textual feature extractor and the text-based domain adaptation, TDAR gains significant improvement. TDAR outperforms the best baseline 178.47\% for the best case and 11.53\% for the worse case on $F_1$-score, and 195.87\% for the best case and 23.95\% for the worse case on NDCG (without special note, we use relative improvement by default). Since our accuracy boost is caused by two parts, we discuss the effectiveness of them separately in Subsections \ref{subsec:RQ2} and \ref{subsec:RQ3}.

\begin{figure}[ht]
	\centering
	\subfigure[Impact of $\eta^t$ and $\lambda^t$]{
		\includegraphics[scale = 0.26]{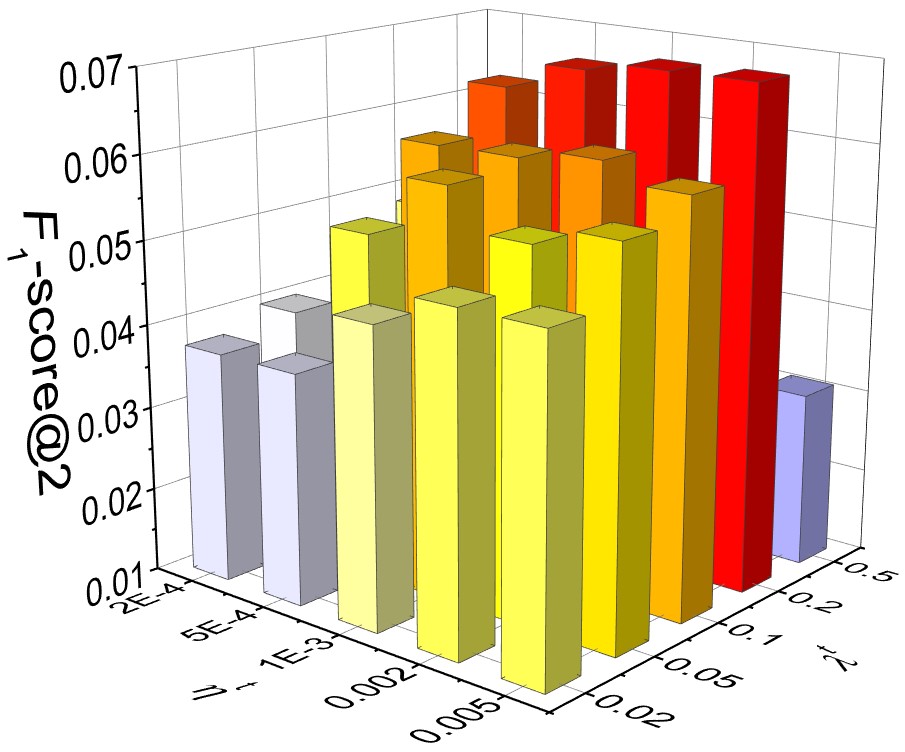}
		\label{subfig:eta_l}
	}
	\hspace{-3mm}
	\subfigure[Impact of $K_3$ ($K_2$)]{
		\includegraphics[scale = 0.21]{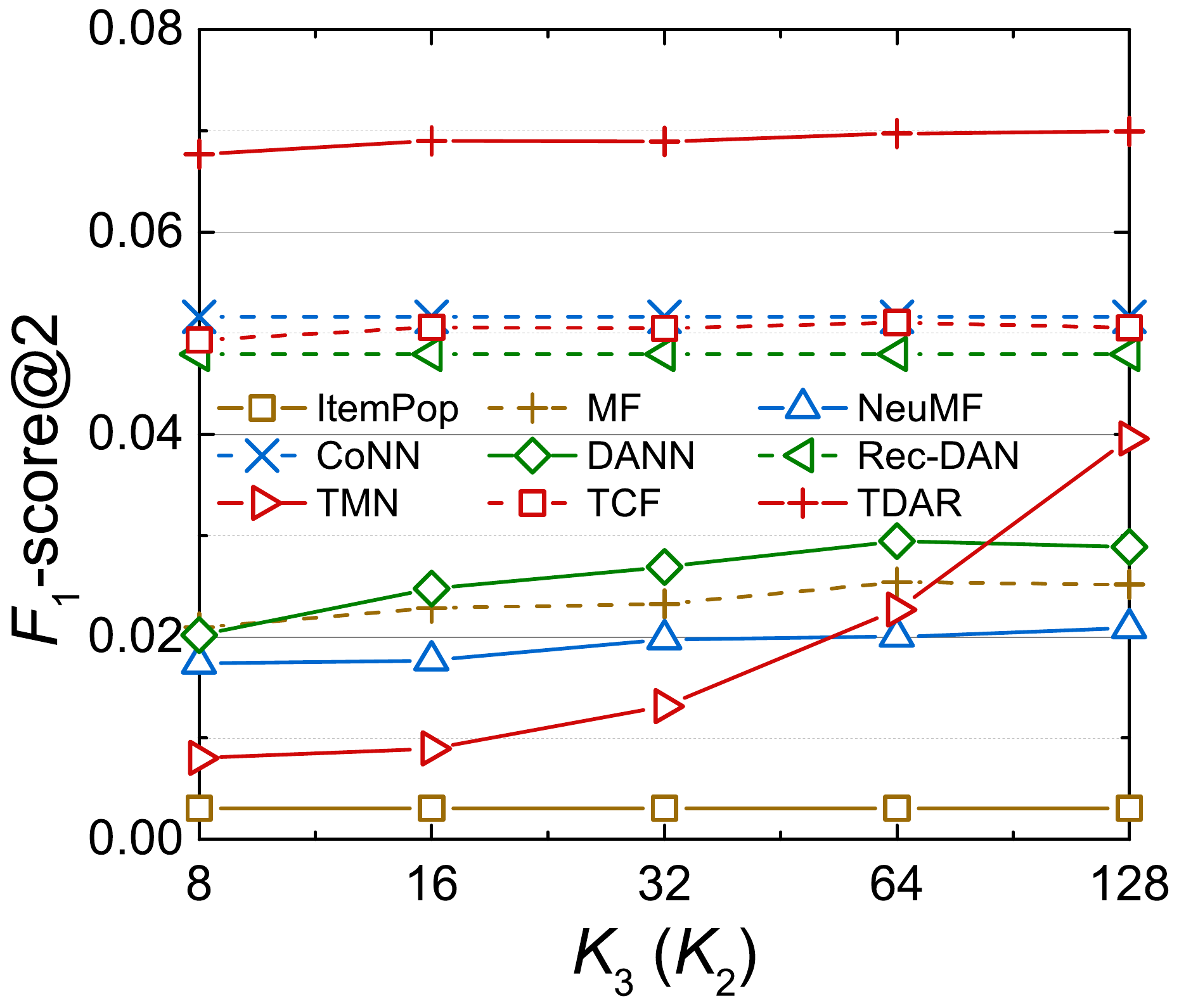}
		\label{subfig:k}
	}
	\vspace{-1mm}
	\caption{Model tuning (\textit{Videos}, validation set)\vspace{-2mm}}
	\label{fig:model_tuning}
\end{figure}

We then report some tuning details in Figure \ref{fig:model_tuning}. To save space, we only report the result on \textit{Videos} due to the similar performance. Figure \ref{subfig:eta_l} shows the impact of learning rate $\eta^t$ and regularization coefficient $\lambda^t$. TDAR achieves the best performance when $\eta^t=0.005$ and $\lambda^t=0.2$. Figure \ref{subfig:k} illustrates the sensitivity analysis of latent factor number (denoted as $K_3$ for latent factor models and $K_2$ for TMN). In TMN, we use semantic features pretrained on GoogleNews corpus by word2vec \cite{word2vec}, and $K_1$ is fixed to 300, thus we tune TMN only with respect to $K_2$. Note that ItemPop, CoNN, and Rec-DAN are not latent factor models, their performance keeps constant. Larger $K_3$ ($K_2$) means stronger representation ability, however, when the sparse dataset cannot provide enough information, strong representation ability helps a little, therefore the performance of MF, NeuMF, and DANN increases very slightly with the increasing of $K_3$. TCF and TDAR is stronger in representation ability -- they make prediction with both textual features and latent factors, hence the increasing of $K_3$ leads to almost no improvement. The supervision of general models is user-item interaction data while of TMN is user-item-word interaction data, which provides much more information, thus the performance of TMN increases significantly with the increasing of $K_2$.

\subsection{Effectiveness of Textual Information (RQ2)}
\label{subsec:RQ2}

In this subsection, we discuss the effectiveness of our textual features. We proposed a memory network TMN as the feature extractor, and injected the feature into MF to propose TCF. The performance of TMN and TCF is also shown in Figures \ref{fig:experiment_result} and \ref{subfig:k}. 

We first compare the text-based models (TMN, TCF, and CoNN) with pure latent factor models (MF and NeuMF). They perform quite differently on the datasets with different sparsity --- the enhancement gained from textual information increases significantly with the increasing of sparsity. On the most sparse dataset \textit{CDs}, performance of MF and NeuMF is extremely bad and CoNN outperforms MF 8.16 times on $F_1$-score for the best case. Compared with \textit{CDs}, \textit{Videos} is much denser, where the performance of text-based models and latent factor models is very similar. Filtered by 30 core, \textit{Movies} is the densest dataset. On \textit{Movies}, latent factor models perform much better than models with only text features (CoNN and TMN). 

We then compare our proposed textual features (TMN) with the existing one (CoNN). Compared with CoNN, TMN is good at sparse cases. On \textit{CDs} and \textit{Clothes} datasets, TMN performs pretty well --- TMN outperforms CoNN 1.25 times on $F_1$-score for the best case. However, on \textit{Movies} and \textit{Videos}, CoNN performs much better --- CoNN outperforms TMN 88.84\% on $F_1$-score for the best case. TMN and CoNN both have specific superiorities: TMN is good at highlighting important keywords while CoNN can model sequential information. As we discussed in Section \ref{sec:TMN}, sequential information is not important in our task. This viewpoint is supported by the experiment --- TMN outperforms CoNN in sparse cases. 


We finally study the effectiveness of combining textual features with latent factors by comparing TMN and TCF. From Figure \ref{fig:experiment_result} we can observe that we gain more improvement on sparse datasets: TCF outperforms TMN significantly on \textit{CDs}, \textit{Clothes}, and \textit{Videos} while they perform pretty similarly on \textit{Movies}. We consider that on sparse datasets, textual features and latent factors complement each other in encoding user preferences, while on dense datasets, implicit feedback data provides enough information thus extra data helps little.

\subsection{Effectiveness of Our Domain Adaptation Strategy (RQ3)}
\label{subsec:RQ3}
In this subsection, we discuss the effectiveness of our text-based domain adaptation strategy by comparing TDAR against DANN and Rec-DAN. We have compared them according to the accuracy in Subsection \ref{subsec:RQ1}. Considering TDAR outperforms DANN and Rec-DAN significantly mainly due to the outstanding effectiveness of our textual features, we use another measurement: improvement over the basic models to evaluate the effectiveness of each domain adaptation approaches. We measure the absolute improvement, i.e., the difference between performances of a cross-domain model and of its basic model. The basic model of DANN is MF, and learning on the target domain is supervised by both positive and negative samples. The basic model of Rec-DAN is LSTM and learning on the target domain is unsupervised. 

\vspace{-2mm}
\begin{figure}[ht]
	\centering
	\subfigure[Impact of domain similarity]{
		\includegraphics[scale = 0.20]{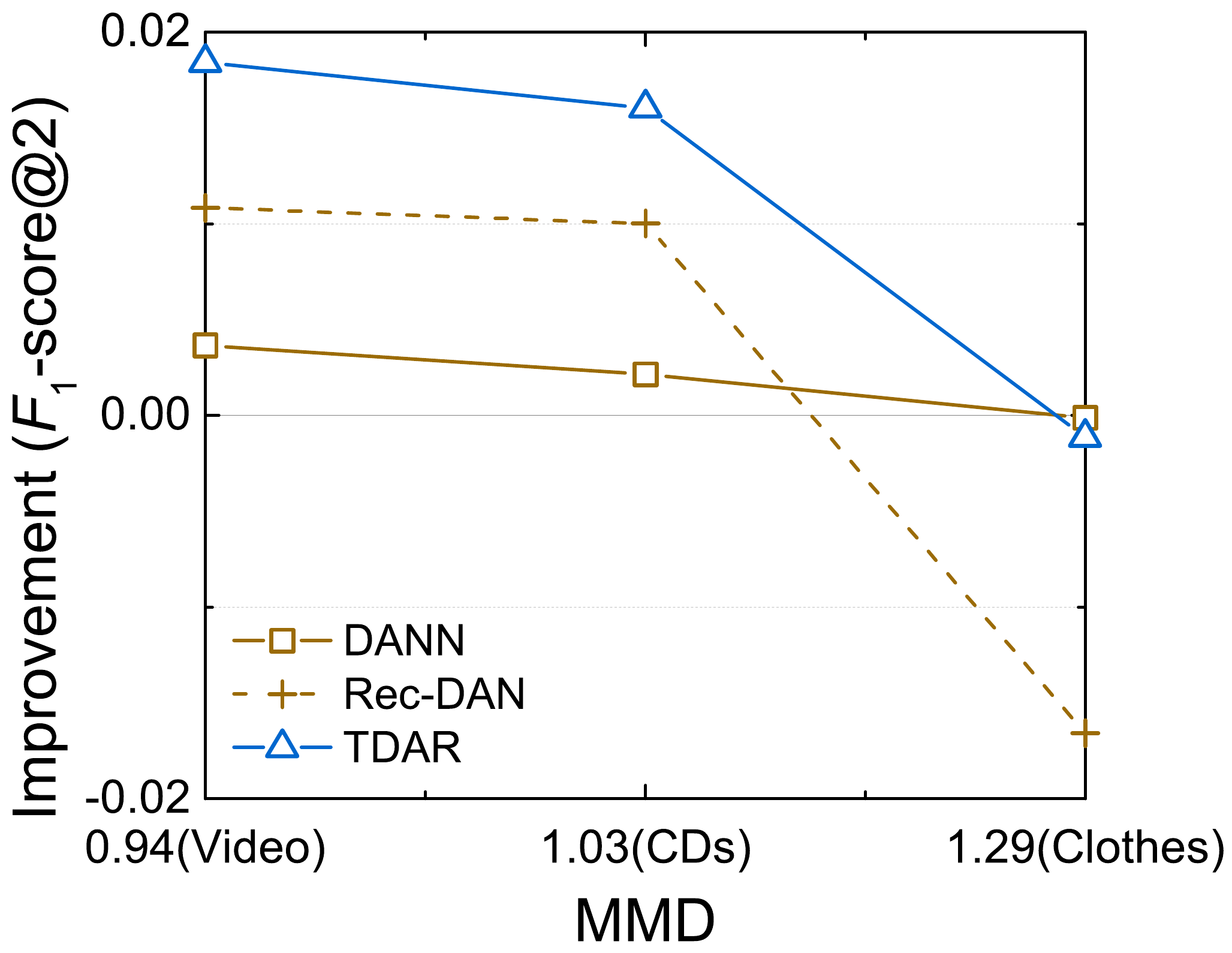}
		\label{subfig:mmd}
	}
	\hspace{-3mm}
	\subfigure[Sampling quality (\textit{Videos})]{
		\includegraphics[scale = 0.20]{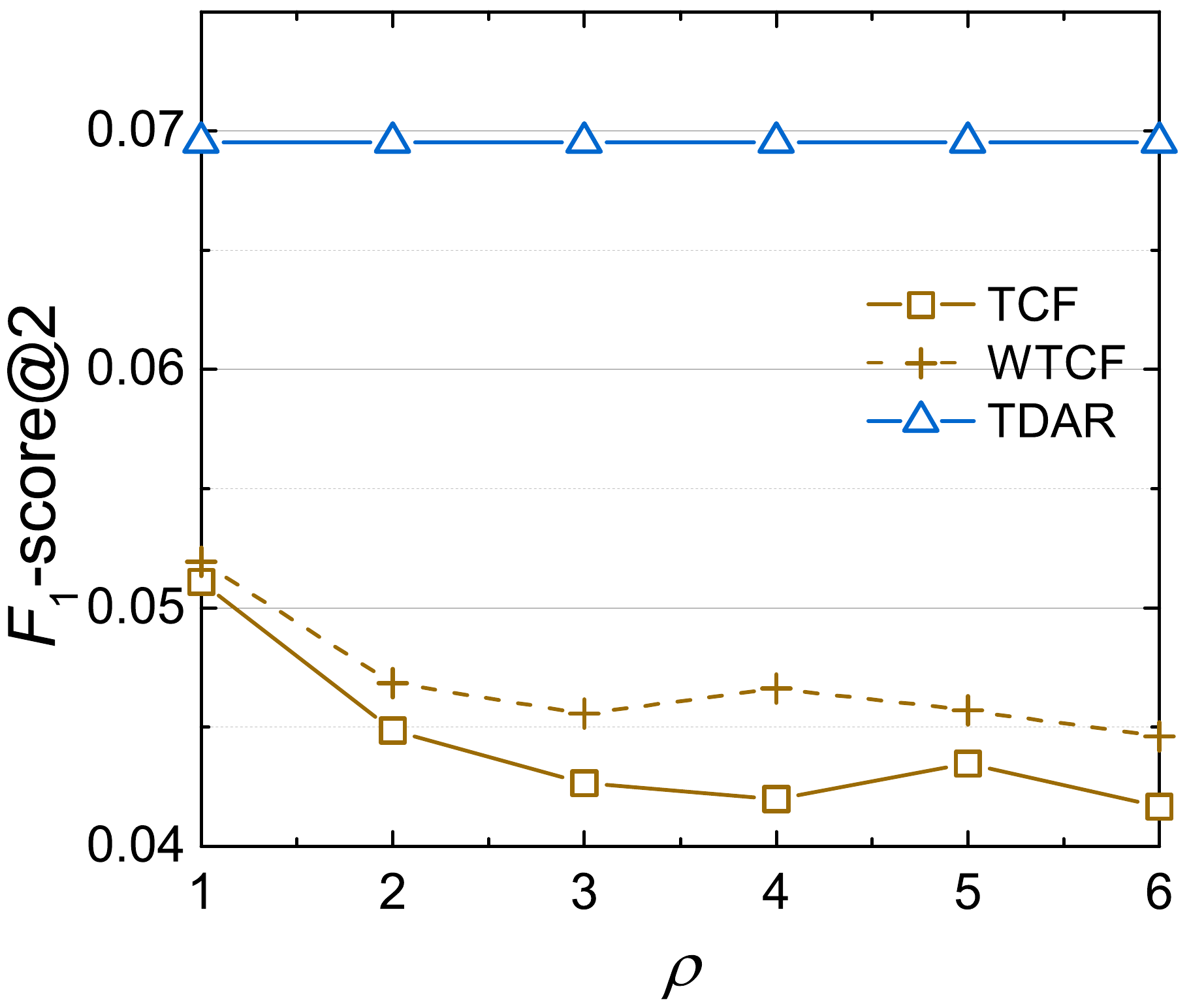}
		\label{subfig:sr}
	}
	\vspace{0mm}
	\caption{Analysis for domain adaptation (validation set)\vspace{0mm}}
	\label{fig:domain_adaptation}
\end{figure}

The improvement of DANN, Rec-DAN, and TDAR on three target datasets is shown in Figure \ref{subfig:mmd} (here we report the result on validation sets, since improvement on test set can be found in Figure \ref{fig:experiment_result}). As shown in Figures \ref{fig:experiment_result} and \ref{subfig:k}, TDAR outperforms cross-domain baselines on predicting accuracy. From Figures \ref{fig:experiment_result} and \ref{subfig:mmd} we can see that TDAR also performs well on accuracy improvement. Figures \ref{subfig:mmd} shows that TDAR outperforms the best cross-domain baseline 70.06\% on improvement for the best case. Comparing Figures \ref{fig:experiment_result} and \ref{subfig:mmd} we can see that another advantage of TDAR is that we can gain stable improvement on test set. For example, on \textit{Videos}, both DANN and TDAR outperform their basic models on the validation set (DANN outperforms MF 9.29\% and TDAR outperforms TCF 35.98\% on $F_1$-score@2), while on the test set, only TDAR still achieves significant improvement (DANN outperforms MF 0.73\% and TDAR outperforms TCF 20.21\% on $F_1$-score@2). This is an experimental evidence for our key idea shown in Figure \ref{fig:embedding_align}: it is not helpful to align embeddings directly in recommendation context. It is also our motivation to propose our text-directed domain adaptation.

As mentioned in Subsection \ref{subsubsec:dataset}, we select three target datasets by considering the different similarity to the source dataset. To quantify the similarity, we calculate Maximum Mean Discrepancy (MMD) \cite{MMD} of text features of source and target datasets. Figure \ref{subfig:mmd} also shows the impact of domain similarity, and MMD in the abscissa is the distance of two distributions. As shown in the figure, the more similar the two datasets are, the more improvement we gain by transfer learning. Since \textit{Movies} and \textit{Clothes} are very different and there is little useful knowledge to transfer, Rec-DAN performs dreadfully in this case since it is unsupervised on the target domain.

As we discussed in Subsection \ref{subsec:TDAR}, we use domain adaptation instead of negative sampling to supervise the model on target domain, therefore our TDAR can be regarded as an approach to improve sampling quality. To evaluate the effectiveness of our sampling strategy, we compare TDAR with two baselines:

\begin{itemize}
	\item{\textbf{TCF:} TCF (Subsection \ref{subsec:TCF}) is the basic model of TDAR optimized by cross entropy (Equation (\ref{equ:loss})). To improve the sampling quality, we randomly select $\rho$ negative samples for each positive one.}
	\item{\textbf{WTCF:} We introduce this \textbf{W}eighted \textbf{TCF} method inspired by Weighted Matrix Factorization (WMF) \cite{WBPR}. When optimized by cross entropy, we weight each positive sample with 1 and weight each negative sample $i$ with $p_i/\bar{p}$, where $\bar{p}$ is the average popularity and $p_i$ is the popularity of $i$. Popular unobserved items are considered less likely to be neglected hence with more confidence to be true negative samples.}
\end{itemize}

Experiment result is illustrated in Figure \ref{subfig:sr}. In sparse data, conventional sampling strategies take little effect. There is much label noise in negative labels and more negative samples means more noise, thus the performance of TCF and WTCF decreases when $\rho$ increases. The improvement of the weighted sampling strategy is very marginal. Compared with conventional sampling strategies, TDAR provides a novel way to avoid negative sampling: We learn a basic model on source domain with label level supervision, and learn another basic model on target domain with both partial label level supervision and embedding level supervision. Comparing the performance of MF on \textit{Movies} and \textit{CDs}, we can see that the performance on target domains is seriously damaged since there is too much label noise. In contrast, label noise on the source domain is totally acceptable. 
As a result, though TDAR is still affected by noise on source domain, we gain considerable improvement by avoiding noise on the target domain.

\section{Conclusion}
\label{sec:conclusion}
In real-world applications, we usually face extremely sparse datasets, where conventional sampling strategies cause much label noise. In this paper, we aim to learn knowledge from dense datasets and transfer it to sparse ones. To do so, we focus on cross-domain recommendation without overlap by aligning embeddings. As the embeddings of different datasets encode different information, aligning them directly is meaningless. To address this gap, we use textual features to direct domain adaptation. We first map users and items of all domains to the same semantic space to construct the textual representation, and then align embeddings with these domain-invariant features. Comprehensive experiments show that our text-enhanced domain adaptation strategy for recommendation outperforms existing approaches significantly. For the future work, we are interested in exploring other side information such as images or tags to direct domain adaptation. We also want to validate the effectiveness of our transfer learning approach in other graph embedding tasks such as social networks.